\documentclass[conference]{IEEEtran}
\IEEEoverridecommandlockouts

\usepackage{cite}
\usepackage{graphicx}
\usepackage{textcomp}
\usepackage{xcolor}
\usepackage[hyphens]{url}
\usepackage[normalem]{ulem}
\usepackage{bbding}
\usepackage{pifont}

\usepackage{amsmath,amssymb,amsfonts, amsthm}
\usepackage{graphicx}
\usepackage{xcolor}
\usepackage[linesnumbered, ruled, vlined]{algorithm2e}
\usepackage{algpseudocode}
\usepackage{multirow}
\usepackage[makeroom]{cancel}
\usepackage{setspace}
\usepackage{enumitem}
\usepackage{textcomp}
\usepackage{extarrows}
\usepackage{hyperref}

\def\BibTeX{{\rm B\kern-.05em{\sc i\kern-.025em b}\kern-.08em
    T\kern-.1667em\lower.7ex\hbox{E}\kern-.125emX}}

\usepackage{titlesec}
\makeatletter
\g@addto@macro{\normalsize}{%
  \setlength{\abovedisplayskip}{3pt plus 0.5pt minus 1pt}
  \setlength{\belowdisplayskip}{3pt plus 0.5pt minus 1pt}
  \setlength{\abovedisplayshortskip}{0pt}
  \setlength{\belowdisplayshortskip}{0pt}
  \setlength{\intextsep}{4pt plus 1pt minus 1pt}
  \setlength{\textfloatsep}{4pt plus 1pt minus 1pt}
  \setlength{\skip\footins}{5pt plus 1pt minus 1pt}
  }
\makeatother

\begin{document}

\pdfpagewidth=8.5in
\pdfpageheight=11in

\newcommand{\squishlist}{
 \begin{list}{$\bullet$}
  { \setlength{\itemsep}{0pt}
     \setlength{\parsep}{2.25pt}
     \setlength{\topsep}{2.75pt}
     \setlength{\partopsep}{0pt}
     \setlength{\leftmargin}{1.5em}
     \setlength{\labelwidth}{1em}
     \setlength{\labelsep}{0.5em} } }
\newcommand{\squishend}{
  \end{list}  }

\newcommand{\bone}{\ding{182}}
\newcommand{\btwo}{\ding{183}}
\newcommand{\bthree}{\ding{184}}
\newcommand{\bfour}{\ding{185}}
\newcommand{\bfive}{\ding{186}}
\newcommand{\bsix}{\ding{187}}
\newcommand{\bseven}{\ding{188}}
\newcommand{\beight}{\ding{189}}
\newcommand{\bnine}{\ding{190}}
\newcommand{\bten}{\ding{191}}
\newcommand{\beleven}{\ding{192}}

\newcommand{\todo}[1]{\textcolor{blue}{#1}}

\newcommand{\TitleAbbr}{Helios}

\newcommand{\placeholder}{{\color{gray}{xxx xx xx x xx x xx xxxx xxxx xxx xx xx x xx x xx xxxx xxxxxxx xx xx x xx x xx xxxx xxxx xxx xx xx x xx x xx xxxx xxxx xxx xx xx x xx x xx xxxx xxxx xxx xx xx x xx x xx xxxx xxxx xxx xx xx x xx x xx xxxx xxxx xxx xx xx x xx x xx xxxx xxxx xxx xx xx x xx x xx xxxx xxxx xxx xx xx x xx x xx xxxx xxxx xxx xx xx x xx x xx xxxx xxxx xxx xx xx x xx x xx xxxx xxxx xxx xx xx x xx x xx xxxx xxxx xxx xx xx x xx x xx xxxx xxxx. }}}

\newcommand{\placeholdershort}{{\color{gray}{xxx xx xx x xx x xx xxxx xxxx xxx xx xx x xx x xx xxxx xxxxxxx xx xx x xx x xx xxxx xxxx xxx xx xx x xx x xx xxxx xxxx xxx xxxx xxxx xxx xx xx x xx x xx xxxx xxxx xxx xx xx x xx x xx xxxx xxxx. }}}


\pagenumbering{arabic}

\title{Hardware-Software Co-design for 3D-DRAM-based LLM Serving Accelerator}
\author{
Cong Li\textsuperscript{1} \quad
Yihan Yin\textsuperscript{1} \quad
Chenhao Xue\textsuperscript{1} \quad
Zhao Wang\textsuperscript{1} \quad
Fujun Bai\textsuperscript{2} \quad
Yixin Guo\textsuperscript{2} \\
Xiping Jiang\textsuperscript{2} \quad
Qiang Wu\textsuperscript{3} \quad
Yuan Xie\textsuperscript{4} \quad
Guangyu Sun\textsuperscript{1} \\[0.2em]

\textsuperscript{1}\textit{Peking University} \quad
\textsuperscript{2}\textit{Xi'an UniIC Semiconductors} \quad
\textsuperscript{3}\textit{Houmo AI} \quad
\textsuperscript{4}\textit{HKUST}
}

\maketitle

\begin{abstract}

Large language models (LLMs) have been widely deployed for online generative services, where numerous LLM instances jointly handle workloads with fluctuating request arrival rates and variable request lengths.
To efficiently execute coexisting compute-intensive and memory-intensive operators, near-memory processing (NMP) based computing paradigm has been extensively proposed.
However, existing NMP designs adopt coarse-grained KV cache management and 
inflexible attention execution flow. Such limitations hinder these proposals from efficiently handling \textit{highly dynamic} LLM serving workloads, limiting their ability to accelerate LLM serving.

To tackle these problems, we propose \uline{H}e\uline{L}io\uline{S}, a \uline{H}ybrid-bonding-based \uline{L}LM \uline{S}erving accelerator.
\TitleAbbr~aims to bridge the fundamental gap between the dynamic nature of KV cache management in LLM serving and the distributed, non-uniform memory abstraction among NMP processing engines (PEs).
To this end, we design both the intra-PE execution flow and the inter-PE communication primitives for distributed tiled attention execution.
We further propose \textit{spatially-aware} KV cache allocation mechanism to balance the attention workload distribution while minimizing the inter-PE data transfer overhead.
Compared with existing GPU/NMP designs, \TitleAbbr~achieves 3.25× (geomean) speedup and 3.36× (geomean) better energy efficiency, along with up to 72\%/76\% P50/P99 time-between-tokens degradation.

\end{abstract}
\section{Introduction}

Large Language Models (LLMs) have enabled many generative tasks, such as code completion~\cite{chen2021evaluating,nijkamp2022codegen}, chatbot~\cite{chatgpt,gemini}, and reasoning~\cite{wei2022emergentabilitieslargelanguage,wei2023chainofthoughtpromptingelicitsreasoning}.
Such capability drives cloud companies to host LLMs as online services~\cite{chatgpt,gemini,claude,deepseek}.
In LLM serving, multiple LLM instances jointly handle highly dynamic request workloads, characterized by widely varying sequence lengths (as high as 10K tokens) and rapidly fluctuating arrival rates (as high as an order of magnitude)~\cite{xiang2025servegen}.
Therefore, each instance conducts inference under both small (e.g., 4-8) and large (e.g., 32-64) batches~\cite{zhong2024distserve}, each comprising requests with non-uniform context lengths~\cite{kwon2023efficient}.

LLM inference contains two stages: prefill and decoding. During the prefill stage, the LLM instance processes the input token sequence (i.e., prompt) in one shot and produces the first output token. 
Then, during the decoding stage, the instance takes the last token as input and generates one new token in each step, incrementally building the full output.
Since LLM's self-attention~\cite{vaswani2017attention} requires each token to interact with all historical tokens, the intermediate vectors associated with each token are persistently stored throughout inference to avoid re-computation, which is known as KV cache~\cite{pope2022efficientlyscalingtransformerinference}.

Given the highly dynamic LLM serving workloads, LLM instances need to process both compute- and memory-intensive operators. For example, operators in the prefill stage are typically compute-intensive, while the small-batch fully-connected (FC) layers and multi-head attention operators (MHA) during the decoding stage
are memory-bound~\cite{park2024attacc}. 
Although centralized processors (e.g., GPU/TPU) offer high compute power to compute-intensive operators, their limited bandwidth (e.g., two orders of magnitude lower~\cite{tpuv4}) leads to poor performance for memory-bound operators.
To mitigate this issue, Near-Memory Processing (NMP) accelerators has been extensively proposed~\cite{heo2024neupims,park2024attacc,yun2024duplex},\cite{lee2021hardware,he2020newton,kim2024breakthrough,lee20221ynm}, which integrates processing engines (PEs) into memory modules (e.g., HBM cubes) to enhance bandwidth provision.
By collaborating with centralized processors, NMP accelerators have achieved superior performance for LLM inference.

However, existing NMP designs lack flexibility in both system and architecture, hindering them from efficiently handling dynamic LLM serving workloads:
\textit{\uline{At the system level}}, attention heads are statically assigned to fixed NMP modules (e.g., NMP-enabled HBM cubes). This coarse-grained scheme not only causes compute load imbalance, but also incurs severe memory fragmentation, reducing accelerator's effective
serving capacity.
\textit{\uline{At the architecture level}}, existing NMP compute logic also requires each attention head's full KV cache to reside within identical NMP modules during attention computation, hindering the intra-sequence KV cache partitioning needed for fine-grained dynamic KV cache management.

To tackle these problems, we propose \TitleAbbr, a hybrid-bonding-based LLM serving accelerator. We aim to develop \textit{NMP-Native} fine-grained attention computation and dynamic KV cache management mechanisms, addressing the fundamental mismatch between LLM serving's \textit{dynamic nature} and NMP's \textit{non-uniform memory abstraction}.
To this end, we customize PE architecture, intra-PE operator execution, and inter-PE communication primitives to enable distributed tiled attention via hybrid bonding, an emerging technology that effectively boosts NMP compute power~\cite{zhao2025insightsdeepseekv3scalingchallenges,yue2024exploiting,li2025h2,pan2025stratum}.
For system scheduling, we propose \textit{spatially-aware} KV cache allocation to balance the attention workload distribution while minimizing the inter-PE data transfer overhead.
To summarize, we have made the following contributions: 
\squishlist
    \item We conduct an in-depth analysis of existing NMP designs' limitations under highly dynamic serving workloads.
    \item We propose \TitleAbbr's architecture, enabling \textit{NMP-Native} distributed tiled attention in end-to-end LLM inference.
    \item We propose \TitleAbbr's system design, enabling \textit{NMP-Native} dynamic KV cache management with full consideration of NMP architecture's non-uniform memory abstraction.
\squishend
Extensive experiments demonstrate that \TitleAbbr~achieves 3.25× (geomean) speedup and 3.36× (geomean) better energy efficiency compared with existing GPU/NMP designs.
For serving performance, \TitleAbbr~achieves up to 72\%/76\% P50/P99 time-between-tokens (TBT) degradation.

\section{Background and Motivation}

\begin{figure}
    \centering
    \includegraphics[width=0.99\columnwidth]{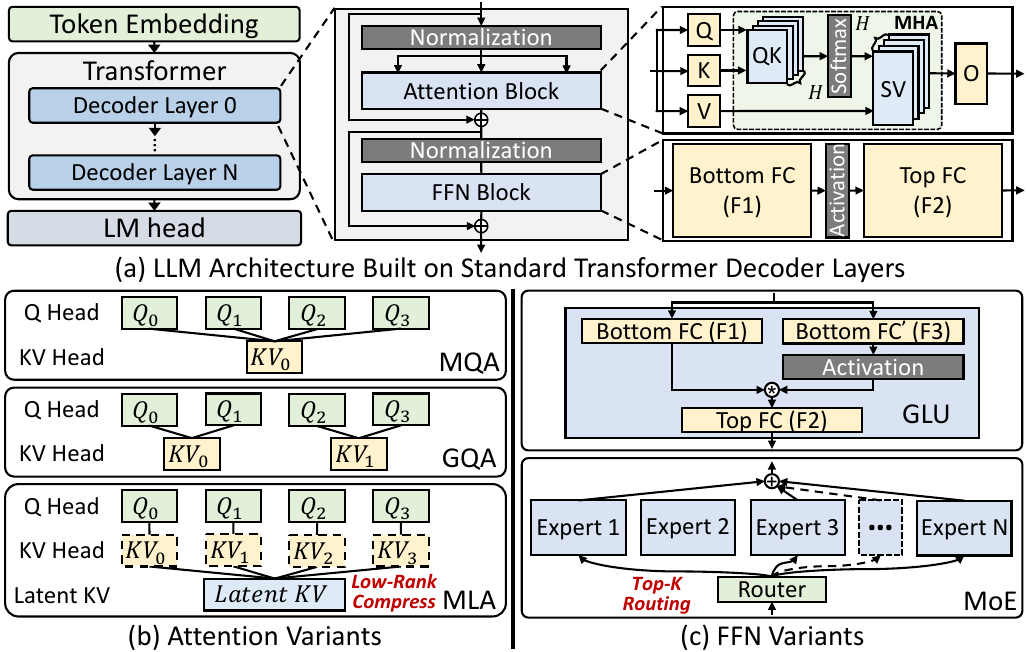}
    \vspace{-1.5em}
    \caption{Standard LLM Architecture and its Variants.}
     \vspace{-0.5em}
    \label{fig:transformer}
\end{figure}

\subsection{Transformer-based LLM Architecture}

As shown in Fig.~\ref{fig:transformer}-(a), mainstream LLMs are built upon the transformer decoder layers~\cite{vaswani2017attention}, which contains an attention block and a feed forward network (FFN) block, along with normalization and residual operators.
In the attention block, the input is first  transformed by three FC layers (Q/K/V) into query/key/value vectors, which are partitioned into $H$ heads (4 in the example) and processed through multi-head attention (MHA).
The result is projected by another FC (O) to produce the block output.
The FFN block contains two FCs.
The output of bottom FC (F1) is processed by the activation function before being processed by the top FC (F2).

Apart from the standard architecture, many variants have been derived.
As shown in Fig.~\ref{fig:transformer}-(b), Multi-query attention (MQA)~\cite{shazeer1911fast} shares one key-value head pair (KV head) across all query heads. Group-query attention (GQA)~\cite{ainslie2023gqa} generalizes it by grouping every $q$ query heads with one KV head (e.g., $q$=2 in Fig.~\ref{fig:transformer}-(b)).
Beyond directly reducing KV head number, multi-latent attention (MLA)~\cite{liu2024deepseek,deepseekai2025deepseekr1incentivizingreasoningcapability,deepseekai2024deepseekv3technicalreport} adds additional FCs to compress all KV heads into one latent head.
For FFN variants, as shown in Fig.~\ref{fig:transformer}-(c), gated linear units (GLUs)~\cite{shazeer2020glu} adds one more bottom FC and applies element-wise multiplication between two bottom FCs' outputs. Mixture of Expert (MoE) block contains a router and multiple expert FFNs. For each token, the router adopts FC and softmax to select top-K experts, whose outputs are combined via weighted summation.

\subsection{LLM Serving}
\label{sec:llm-serving}

\begin{figure}
    \centering
    \includegraphics[width=0.99\columnwidth]{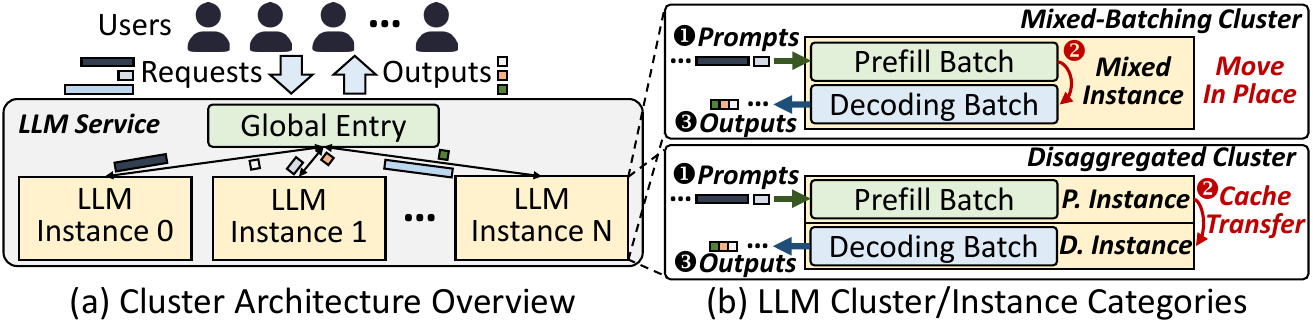}
    \vspace{-1.75em}
    \caption{LLM Serving Cluster Overview.}
     \vspace{-1.25em}
    \label{fig:serving}
\end{figure}

LLM has been widely deployed in online services~\cite{chatgpt,gemini,deepseek}.
As shown in Fig.~\ref{fig:serving}-(a), LLM service receives requests from multiple users, dispatches them to a cluster hosting numerous LLM instances, and returns decoding outputs to users.
Since LLM inference contains prefill and decoding stages, there are two instance deployment modes, each aligned with one cluster architecture in Fig.~\ref{fig:serving}-(b):
\textit{\uline{(1) Mixed-batching Cluster}}:
Each instance maintains both prefill and decoding batches, which are jointly processed in each iteration.
Each request is first added to the prefill batch (\bone).
After prompt processing, it is moved to the decoding batch in place (\textcolor{red}{\btwo}), reusing local KV cache for decoding (\bthree).
\textit{\uline{(2) Disaggregated Cluster}}: Instances are specialized for either prefill or decoding. 
Each request is assigned to a prefill-decoding instance pair and is first processed by the prefill instance (\bone).
Then, the decoding instance receives KV cache (\textcolor{red}{\btwo}) and continue decoding (\bthree).
The cache copy overhead can be overlapped by layer-wise KV cache transfer~\cite{qin2025mooncake,patel2024splitwise}.
By avoiding prefill-decoding interference~\cite{zhong2024distserve,patel2024splitwise}, disaggregated cluster achieves superior performance to mixed-batching cluster, which has been validated in large-scale production deployments~\cite{qin2025mooncake}.

LLM serving workloads exhibit \textit{\textbf{high dynamicity}},  which is manifested in two key aspects:
\textit{\uline{(1) Request Arrival}}: LLM services have significant traffic fluctuations. 
Based on production trace analyses~\cite{wang2024towards,xiang2025servegen}, peak-hour loads can exceed off-peak periods by 1-2 orders of magnitude.
Therefore, it is crucial for LLM instances to efficiently handle both small batches at low load and large batches at high load~\cite{zhong2024distserve}.
\textit{\uline{(2) Request Length}}:
As LLMs' context windows continue to expand, token counts vary widely across requests.
Production service analysis~\cite{xiang2025servegen} shows that the max length can be an order of magnitude greater than the average length (e.g., \textasciitilde10K vs. \textasciitilde1K).

\begin{figure}
    \centering
    \includegraphics[width=0.99\columnwidth]{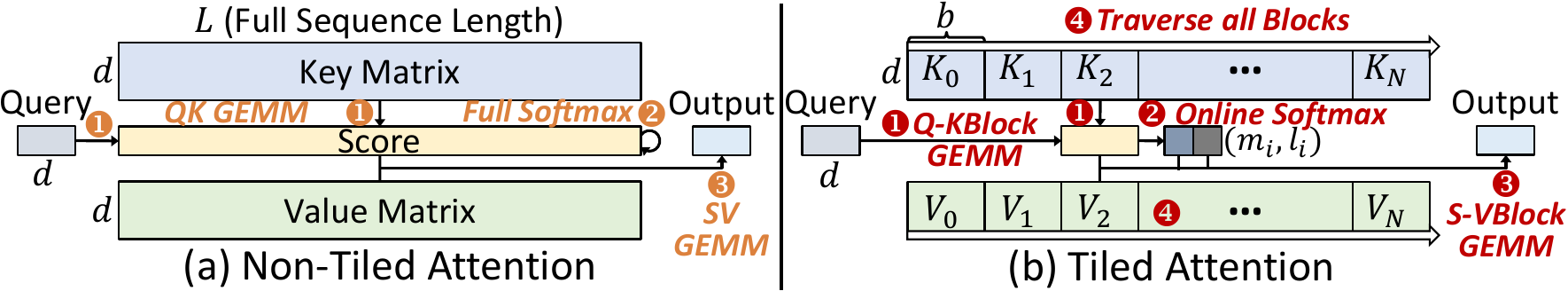}
    \vspace{-1.5em}
    \caption{Illustration of Non-Tiled \& Tiled Attention.}
     \vspace{-0.5em}
    \label{fig:tiled-attention}
\end{figure}

Such dynamicity causes the KV cache maintained in each LLM instance to continuously evolve as serving progresses due to the following reasons:
\uline{(1)} The residency of the KV cache is tied to the lifetime of each request. As requests dynamically enter (e.g., new request arrives) or exit (e.g., request finishes decoding) an LLM instance under varying request rates, the allocated KV cache correspondingly expands or shrinks.
\uline{(2)} KV cache memory footprint and attention computation cost both scale proportionally with request length~\cite{kwon2023efficient,zhong2024distserve}. Therefore, variable request lengths directly affect per-instance KV cache memory usage and attention overhead.
To align with this behavior, dynamic KV cache management~\cite{kwon2023efficient} and fine-grained attention computation algorithms~\cite{dao2022flashattention,dao2023flashattention,shah2024flashattention3fastaccurateattention,flashdecoding} have been proposed, which are widely employed in state-of-the-art serving systems such as vLLM~\cite{kwon2023efficient} and SGLang~\cite{sglang}:
\textit{\uline{(1) Blockwise KV Cache Management}}:
As shown in Fig.~\ref{fig:tiled-attention}-(b), it partitions each request’s full KV cache into blocks of $b$ tokens. By dynamically allocating free block slots in accelerator memory, it significantly improves the memory utilization.
\textit{\uline{(2) Tiled Attention}}:
To align with blockwise management mechanism, online softmax~\cite{milakov2018onlinenormalizercalculationsoftmax} is introduced to enable block partition during attention computation.
Tiled attention has two versions: iterative and reduction. 
Fig.~\ref{fig:tiled-attention}-(b) shows iterative tiled attention's execution flow, and Algorithm~\ref{algo:tiled-attention} summarizes its formulae.
Without loss of generality, these formulae are exemplified using one MHA head, which are applicable to any attention variant.
For each KV block, iterative tiled attention performs GEMM between query and key block (\textcolor{red}{\bone}), computes online softmax and updates intermediate results (\textcolor{red}{\btwo}), conducts GEMM between score block and value block, and accumulate partial sums (\textcolor{red}{\bthree}).
After traversing all blocks (\textcolor{red}{\bfour}), the complete output is obtained.
Alternatively, pre-computed partial results from different blocks can be combined via tiled attention's reduction version:
Assume two blocks' intermediate results are ($\boldsymbol{O}_1,m_1,l_1$) and ($\boldsymbol{O}_2,m_2,l_2$), using the notation in Algorithm~\ref{algo:tiled-attention}.
The attention output is given by:
\begin{equation}
\label{eq:fd-factor-gen}
\textstyle{m=max(m_1,m_2),\  
e_1 = l_1 \cdot e^{m_1-m},\ 
e_2 = l_2 \cdot e^{m_2-m}}
\end{equation}
\begin{equation}
\label{eq:fd-sum}
l = e_1 + e_2,\ 
\boldsymbol{O}=\tfrac{e_1}{l} \cdot \boldsymbol{O}_1 + \tfrac{e_2}{l} \cdot \boldsymbol{O}_2
\end{equation}

\begin{algorithm}[t]
\setstretch{1.0}
\footnotesize
\caption{\small Iterative Tiled Attention (MHA Example)}
\label{algo:tiled-attention}

\textbf{Input:} Query $\boldsymbol{Q} \in \mathbb{R}^{1 \times d}$, KV blocks $\{(\boldsymbol{K}_0,\boldsymbol{V}_0),...,(\boldsymbol{K}_{N},\boldsymbol{V}_{N})\}$, $\boldsymbol{K}_i, \boldsymbol{V}_i \in \mathbb{R}^{b \times d}$ ($d$: head dim, $b$: token number per block).

\textbf{Output:} Attention output $\boldsymbol{O} \in \mathbb{R}^{1 \times d}$.

$\boldsymbol{O} \leftarrow \vec{0}$, $m_{-1} = -\infty$, $l_{-1} = 0$

\For(){$i=0$ to $i=N$}{

$\boldsymbol{x}_i \leftarrow \boldsymbol{Q} \times \boldsymbol{K}_i^{\top}$ \textcolor{blue}{// GEMM between query and key block}

$m_i \leftarrow max(m_{i-1}, max(\boldsymbol{x}_i))$  \textcolor{blue}{// Online softmax begins} 

$\boldsymbol{e}_i \leftarrow e^{\boldsymbol{x}_i-m_i}$

$l_i \leftarrow l_{i-1} \cdot e^{m_{i-1}-m_i} + sum(\boldsymbol{e}_i)$

$\alpha \leftarrow \frac{l_{i-1} \cdot e^{m_{i-1}-m_i}}{l_i}$, $\boldsymbol{p}_i \leftarrow \frac{\boldsymbol{e}_i}{l_i}$  \textcolor{blue}{// Online softmax ends}

$\boldsymbol{O}_i \leftarrow \boldsymbol{p}_i \times \boldsymbol{V}_i$  \textcolor{blue}{// GEMM between score block and value block}

$\boldsymbol{O} \leftarrow \alpha \cdot \boldsymbol{O} + \boldsymbol{O}_i$  \textcolor{blue}{// Ouptut accumulation}

}

\Return $\boldsymbol{O}$

\end{algorithm}

\subsection{NMP-empowered LLM Accelerators}

To efficiently handle coexisting compute-intensive and memory-intensive operators, NMP-based LLM accelerators have been widely proposed~\cite{heo2024neupims,kim2024sk,kim2023samsung,li2024specpim,park2024attacc,yun2024duplex,li2025h2,he2025papi,gu2025pim,pan2025stratum}.
Apart from equipping centralized processors (e.g., GPU/TPU) for compute-intensive operators, they also place processing engines (PEs) into DRAM modules to boost bandwidth for memory-bound operators.
For example, in edge scenarios, Samsung/SK-Hynix propose LPDDR-PIM products for
on-device LLM inference~\cite{kim2023samsung,kim2024sk}.
H$^2$-LLM adopts hybrid bonding to enhance NMP compute power for low-batch inference~\cite{li2025h2}.
As to LLM serving, AttAcc/NeuPIMs target MHA by placing PEs into HBM's DRAM dies~\cite{park2024attacc,heo2024neupims}.
PAPI extends this approach to low-batch FC operators~\cite{he2025papi}.
Duplex adds extra Through Silicon Vias (TSVs) and places PEs on HBM cube's buffer die for GQA and MoE's FC operators~\cite{yun2024duplex}.
Stratum adopts monolithic 3D DRAM to further enhance the inference performance of MoE model's operators~\cite{pan2025stratum}.

\subsection{Limitations of Serving-Oriented NMP Designs}
\label{sec:nmp-limitation}

\begin{figure}
    \centering
    \includegraphics[width=0.99\columnwidth]{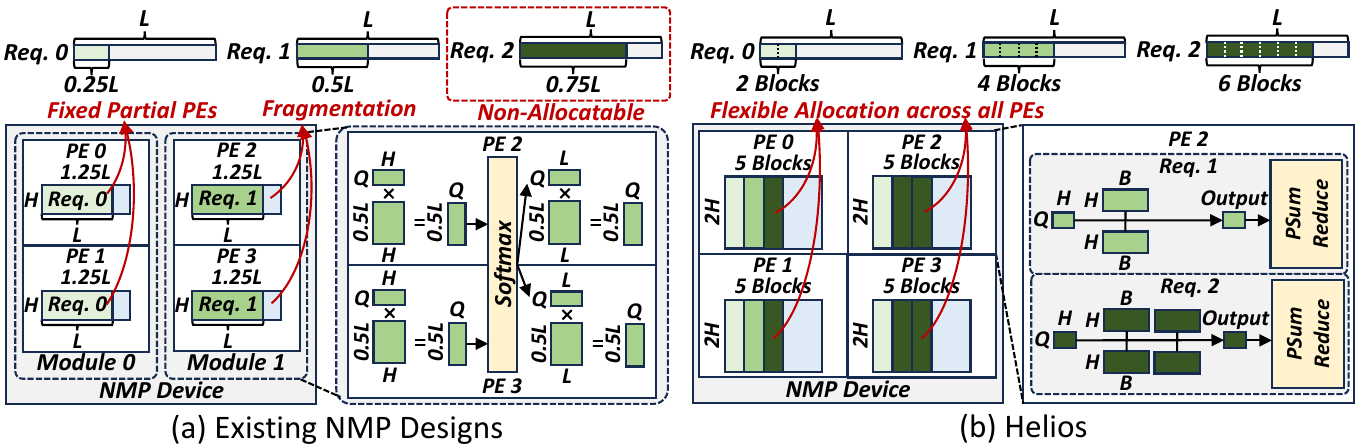}
    \vspace{-1.5em}
    \caption{KV Cache Management and Attention Execution Comparison.}
     \vspace{-0.5em}
    \label{fig:nmp-comparison}
\end{figure}

Although existing LLM-serving-oriented NMP designshave reported better performance than centralized processors, they lack support for dynamic KV cache management and fine-grained attention computation mechanisms, limiting their efficiency for highly dynamic serving workloads:

\noindent\textbf{Limitation \#1: Coarse-grained KV cache management.}
As summarized in Table~\ref{tab:nmp-baseline}, existing NMP designs group PEs into NMP modules corresponding to memory module organization (e.g., HBM cubes/channels), and \textbf{\textit{statically}} allocate the full-length KV cache of each attention head to a fixed NMP module throughout the request's lifetime.
For example, in Fig.~\ref{fig:nmp-comparison}-(a), each request is placed on two PEs within the same NMP module during the entire execution.
This coarse-grained scheme leads to resource inefficiency from two aspects:

\begin{table}[t]
\centering
\caption{Comparison of Existing NMP-based LLM Serving Accelerators}
\label{tab:nmp-baseline}

\vspace{-0.5em}

\resizebox{0.475\textwidth}{!}{

\begin{tabular}{|c|c|c|c|c|}

\hline
\multirow{3}{*}{Name} 
  & \multicolumn{2}{c|}{KV Cache Management} 
  & \multicolumn{2}{c|}{Attention Computation} \\ \cline{2-5}
  & Flexibility & \begin{tabular}[c]{@{}c@{}} Allocation Scope\\ (per KV Head) \end{tabular} & \begin{tabular}[c]{@{}c@{}} Execution\\ Flow \end{tabular} & \begin{tabular}[c]{@{}c@{}} Execution Scope\\ (per KV Head) \end{tabular} \\ 
\hline

AttAcc~\cite{park2024attacc}  & \multirow{7}{*}{\begin{tabular}[c]{@{}c@{}} Full-length\\ Static \end{tabular}}  & One HBM Cube & \multirow{7}{*}{\begin{tabular}[c]{@{}c@{}} Localized\\ Non-tiled \end{tabular}}  & One HBM Cube \\ 

\cline{1-1}\cline{3-3}\cline{5-5}

NeuPIMs~\cite{heo2024neupims}   &    & One HBM Channel &   & One HBM Channel\\ 

\cline{1-1}\cline{3-3}\cline{5-5}

Duplex~\cite{yun2024duplex}   &    & One HBM Cube &   & One HBM Cube  \\ 

\cline{1-1}\cline{3-3}\cline{5-5}

PAPI~\cite{he2025papi}   &    & One HBM Cube &   & One HBM Cube  \\ 

\cline{1-1}\cline{3-3}\cline{5-5}

CENT~\cite{gu2025pim}   &    & \begin{tabular}[c]{@{}c@{}} GDDR Channels\\ (Equal to TP size) \end{tabular} &  & \begin{tabular}[c]{@{}c@{}}  GDDR Channels\\ (Equal to TP size) \end{tabular} \\

\cline{1-1}\cline{3-3}\cline{5-5}

Stratum~\cite{pan2025stratum}   &  & Four NMP PEs &  & Four NMP PEs  \\ 

\hline

\textbf{\TitleAbbr~(ours)}  & \begin{tabular}[c]{@{}c@{}} \textbf{Blockwise}\\ \textbf{Dynamic} \end{tabular}  & \begin{tabular}[c]{@{}c@{}} \textbf{All NMP PEs} \end{tabular} & \begin{tabular}[c]{@{}c@{}} \textbf{Distributed}\\ \textbf{Tiled} \end{tabular} & \begin{tabular}[c]{@{}c@{}} \textbf{All NMP PEs} \end{tabular} \\

\hline

\end{tabular}

}

\vspace{-1.em}

\end{table}

\uline{\textit{(1) Compute Load Imbalance}}. 
To fully utilize all NMP modules, the above strategy requires all ongoing requests' KV head number to be no less than NMP module count.
However, as attention variants notably reduce LLM's KV head number, it is difficult to fully utilize all modules under highly dynamic arrival rates.
For example, consider serving LLaMA3 70B~\cite{dubey2024llama} (8 KV heads) on 8 A100 GPUs under tensor parallelism (TP) of 8.
If all five HBM cubes per GPU are NMP cubes, full resource utilization demands at least five requests.
However, such concurrency cannot be satisfied at low loads~\cite{xiang2025servegen}, resulting in idle NMP resources.
Additionally, decoding attention's compute cost scales linearly with request length~\cite{zhong2024distserve}.
Given that request length varies by tens of thousands of tokens~\cite{xiang2025servegen}, assigning each KV head's all tokens as a whole makes it difficult to ensure load balance across NMP modules.

\uline{\textit{(2) Memory Fragmentation}}. 
On one hand, due to NMP module's capacity limitation, it is possible that no single module can accommodate an entire request, even when the total remaining capacity is sufficient. This inter-module fragmentation degrades the effective serving capacity~\cite{sun2024llumnix}.
On the other hand, as request length grows during decoding, pinning KV heads to fixed NMP modules can lead to capacity overflow.
This issue can be mitigated by preserving token slots based on LLM's max context window. However, since requests rarely fill the full window~\cite{xiang2025servegen}, this approach incurs severe memory capacity waste, hurting the serving capability.
In Fig.~\ref{fig:nmp-comparison}-(a)'s example, assume a max context length of $L$, key/value vector length $H$, and per-PE KV cache capacity of ($1.25L,H$).
Although requests 0 and 1 have actual lengths of $0.25L$ and $0.5L$, existing NMP designs still reserve KV cache for length $L$, leading to $\frac{0.75L+0.5L}{L+L}$=62.5\% wasted space.
Moreover, while the total remaining capacity across the four PEs could accommodate another KV cache of length $L$, no individual PE  group has sufficient space. Therefore, request 2 cannot be executed immediately and must wait until request 0 or 1 finishes decoding and releases its allocated space.

\noindent\textbf{Limitation \#2: Inflexible attention execution flow.}
Existing NMP architectures execute attention as two standalone GEMMs~\cite{park2024attacc,kim2023samsung,kim2024sk,he2025papi,heo2024neupims,yun2024duplex,gu2025pim,pan2025stratum}. 
Specifically, for each KV head, its complete key and value matrices are \textbf{\textit{statically}} partitioned across the PEs within a fixed NMP module.
These PEs perform attention computation following the execution flow 
shown in Fig.~\ref{fig:tiled-attention}-(a):
They first collectively compute GEMM between query and the full key matrix (\textcolor{orange}{\bone}). 
Then, the intermediate results undergo the softmax operation to yield the attention scores of the full sequence (\textcolor{orange}{\btwo}).
The scores are then distributed across these PEs according to value matrix's partition.
Finally, these PEs perform GEMM between attention scores and the value matrix to produce the final results (\textcolor{orange}{\bthree}).
In Fig.~\ref{fig:nmp-comparison}-(a)'s example, key and value matrices are partitioned across the two PEs in each group along the context-length $L$ and vector-length $H$, respectively.
With $Q$ query heads, each PE executes two GEMMs: ($Q,H$)×($H,0.5L$) and ($Q,L$)×($L,0.5L$).
Such a static partitioning enforces a fixed context length, fundamentally constraining KV cache dynamism and placement flexibility at the architectural level.
Although this strategy aligns with coarse-grained KV cache management, they jointly limit existing NMP architectures from supporting dynamic KV cache management, degrading their efficiency in LLM serving.

\noindent\textbf{Limitation Analysis:}
The root cause of the above limitations lies in the fundamental gap between the non-uniform memory abstraction inherent in NMP architecture and existing GPU-based dynamic KV cache management algorithm designs.
\uline{\textit{For NMP architecture}}, to fully exploit DRAM bandwidth, it distributes PEs among discrete memory modules (e.g., HBM cubes).
Accordingly, each PE can only directly access its local memory, rather than the accelerator's arbitrary memory modules.
Therefore, to align with such architectural discreteness, existing NMP designs statically assign KV heads to fixed NMP modules.
On the other hand, \uline{\textit{for KV-cache management}}, existing GPU-based algorithms~\cite{kwon2023efficient,dao2022flashattention,dao2023flashattention,shah2024flashattention3fastaccurateattention,flashdecoding} are designed under the assumption that the compute logic has unrestricted access to the accelerator's all memory modules.
However, in NMP architectures, dynamic KV cache management inevitably involves partitioning KV cache blocks across PE arrays that cannot directly access each other’s memory modules.
Designing KV cache management and attention computation strategies that effectively account for such architectural discreteness remains an unresolved challenge for NMP architectures.

\noindent\textbf{Design Goal:}
To address these limitations, we propose \TitleAbbr~to enable \uline{\textit{NMP-native}} dynamic KV cache management.
By dividing KV cache into fine-grained blocks (e.g., block size of $\frac{L}{8}$ in Fig.~\ref{fig:nmp-comparison}-(b)) and distributing them across all PEs, we can balance computation across PEs, eliminate redundant allocation in each request, and fully utilize all PEs' idle capacity.
In addition, by optimizing attention execution around blockwise KV cache management, we can better exploit NMP device's compute capability.
In the following sections, we will elaborate on \TitleAbbr's full-stack design \textemdash~from architecture, to operator execution, and \textit{spatially-aware} KV cache allocation.

\section{\TitleAbbr~Architecture}

\begin{figure}
    \centering
    \includegraphics[width=0.99\columnwidth]{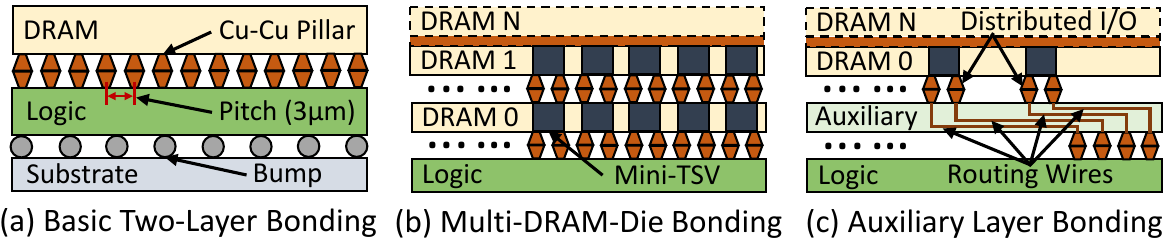}
    \vspace{-1.5em}
    \caption{Basic and Advanced Hybrid Bonding Integration.}
     \vspace{-0.5em}
    \label{fig:hybrid-bonding}
\end{figure}

\begin{figure*}
    \centering
    \includegraphics[width=0.99\linewidth]{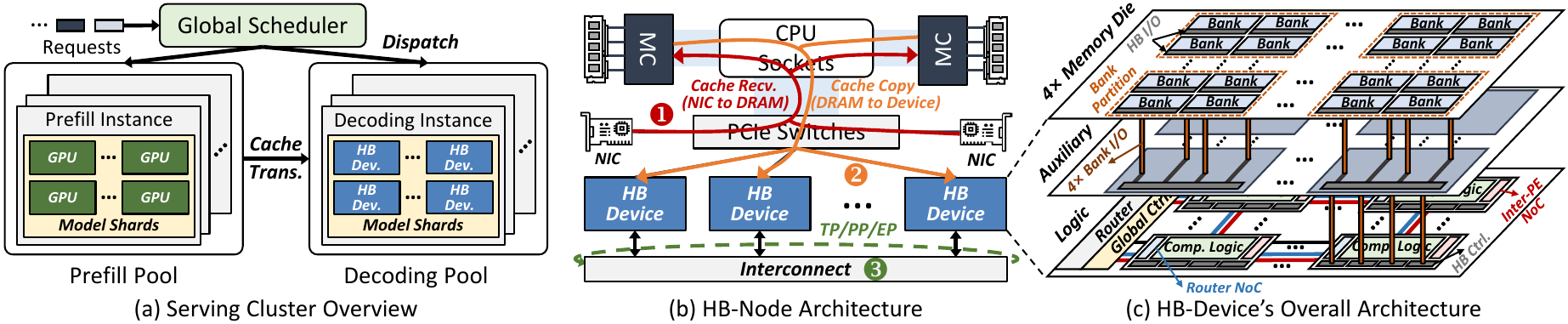} 
    \vspace{-0.5em}
    \caption{Architecture Overview of \TitleAbbr.}
    \vspace{-1.75em}
    \label{fig:architecture}
\end{figure*}

\subsection{The Basis and Advancement of Hybrid Bonding}

\noindent\textbf{Basis:} Before delving into the design details of \TitleAbbr, we first provide an overview of Hybrid Bonding (HB), an integration technology emerging in recent years~\cite{fujun2020stacked,niu2022184qps,wuu20223d,yue2024exploiting,wang2023135,wang20253d,li2025h2}. As shown in Fig.~\ref{fig:hybrid-bonding}-(a), it 3D-stacks the DRAM die onto the logic die via Cu-Cu direct fusion bond.
By offering high I/O density (110,000/mm$^2$) with 3um pitch~\cite{fujun2020stacked,wang2023135}, HB delivers abundant bandwidth for memory-intensive workloads.
Besides, the reduced parasitic capacitance and the low resistance ($<$0.5$\Omega$) enable HB to achieve better memory access energy efficiency than HBM and GDDR memory, reaching as low as 0.66pJ/bit~\cite{wang2023135}.
Moreover, the holistic logic die can be customized to boost the
compute capacity~\cite{yue2024exploiting,li2025h2}, overcoming the compute bottleneck introduced by integrating PEs into DRAM devices (i.e., in-die NMP designs~\cite{li2025h2}).

\noindent\textbf{Advancement:}
In recent years,
more versatile multi-layer bonding
have become feasible~\cite{wang2023135,wang20253d}.
As shown in Fig.~\ref{fig:hybrid-bonding}-(b), beyond stacking a single DRAM die, HB supports multi-layer DRAM bonding by employing mini-TSVs to connect adjacent DRAM dies.
This approach boosts both memory capacity and bandwidth in proportion to DRAM die number.
According to our industry collaborators, current mature process supports four-layer DRAM stacking, providing a capacity comparable to that of mainstream GPUs ($\ge$80GB).

Apart from expanding DRAM capacity, multi-layer bonding can also be used to optimize placement and routing.
To provide sufficient bandwidth, HB partitions the DRAM die into numerous banks with independent I/Os (e.g., 128-256 banks~\cite{yue2024exploiting,li2025h2}).
Directly connecting DRAM I/Os to the logic die as in Fig.~\ref{fig:hybrid-bonding}-(a) would fragment the logic die into many small regions, increasing placement and routing complexity for compute logic. 
To mitigate this issue, an auxiliary logic die can be inserted between the bottom DRAM die and the main logic die to aggregate I/Os from different banks, as shown in Fig.~\ref{fig:hybrid-bonding}-(c). According to our industry collaborators, this approach has negligible impact on bandwidth provision.


\begin{figure}
    \centering
    \includegraphics[width=0.99\columnwidth]{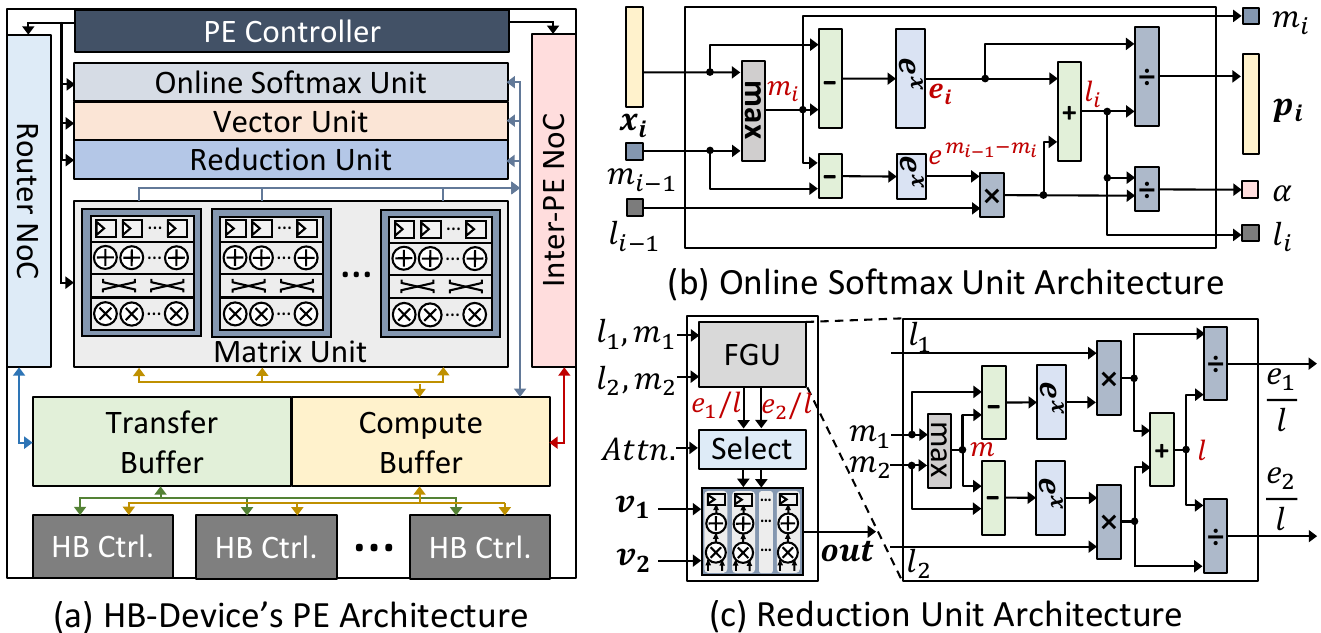}
    \vspace{-1.5em}
    \caption{HB-Device's PE Architecture Design.}
     \vspace{-0.5em}
    \label{fig:pe-arch}
\end{figure}

\subsection{Architecture Overview}

\noindent\textbf{Cluster Architecture:}
\TitleAbbr~employs inter-device heterogeneous architecture, which is fully compatible with state-of-the-art prefill-decoding disaggregation cluster architecture~\cite{patel2024splitwise,qin2025mooncake,zhong2024distserve}.
As shown in Fig.~\ref{fig:architecture}-(a), compute-intensive prefill instances are deployed on GPUs, while decoding instances with memory-bound operators
are assigned to server nodes with hybrid bonding devices (HB-Devices).
Decoding instances receive KV cache from prefill ones, which is coordinated by global scheduler's instance pair allocation.

Note that prefill instances could also be replaced with compute-optimized HB-Devices by adopting more advanced technology nodes to the logic die.
Since GPUs can provide sufficient prefill acceleration, we focus on applying HB to the decoding stage to fully leverage its bandwidth advantages.
Prefill HB-Device design is left as our future work.

\noindent\textbf{Server Node Architecture:}
As shown in Fig.~\ref{fig:architecture}-(b), HB-Devices 
can be integrated into mainstream server architecture~\cite{qin2025mooncake}.
Each device connects to CPU sockets via PCIe switches.
When HB-Node receives new prompt KV cache, it is first delivered to CPU memory through NICs for persistent storage (\textcolor{red}{\bone}) and then copied to HB-Devices via PCIe (\textcolor{orange}{\btwo}).
This process is compatible with high-performance KV cache transfer engines (e.g., MoonCake Transfer Engine~\cite{mooncacke-transfer-engine}).
In this way, prefill instances can stream KV cache layerwise, consistent with industrial best practice~\cite{qin2025mooncake,patel2024splitwise}.
Besides, all devices are connected through high-speed links (NVLink, CXL, etc.)~\cite{ualink,cxl-doc,nvlink} for inter-device tensor/pipeline /expert parallelism (TP/PP/EP) data transfer (\textcolor[rgb]{0.329, 0.510, 0.208}{\bthree}).

\noindent\textbf{Device Architecture:}
As shown in Fig.~\ref{fig:architecture}-(c), each HB-Device stacks four memory dies and one auxiliary die on top of the logic die.
Each memory die contains 256 banks arranged in a $16 \times 16$ grid, which are further grouped into bank partitions with the shape of $n \times n$ (e.g., $n=2$ in Fig.~\ref{fig:architecture}-(c)).
Bank partitions at each DRAM die's same position correspond to the aligned processing engine (PE) on the logic die.
Their I/Os are connected to the auxiliary die using the scheme in Fig.~\ref{fig:hybrid-bonding}-(b). These I/Os are then aggregated using the scheme in Fig.~\ref{fig:hybrid-bonding}-(c).
In this way, HB controllers are consolidated at the bottom of each PE, simplifying logic die's placement and routing compared with distributed I/Os.

The logic die integrates a device router connected to external interconnects, a global controller managing the entire device, and an $m \times m$ PE array ($m=\frac{16}{n}$). 
Each PE can directly access local memory through its HB controllers, and contains two NoC routers for inter-PE communication.
The router NoC links each PE to the device router, handling the transfer of prompt KV cache and TP/PP/EP data.
The inter-PE NoC manages partial sum exchanges during operator execution.
Both NoCs use mesh topologies for PE interconnection and operate on independent data paths to avoid interference.

\noindent\textbf{PE Architecture:}
As shown in Fig.~\ref{fig:pe-arch}-(a), each PE contains a matrix unit, an online softmax unit, a vector unit, and a reduction unit for operator execution.
The matrix unit, built from MAC arrays, performs GEMMs in FC and tiled attention operators. Given the workload and hardware configuration, we can adopt existing performance models~\cite{kwon2020maestro,parashar2019timeloop,lu2021tenet} to determine GEMM's optimal tiling strategies.
Fig.~\ref{fig:pe-arch}-(b) depicts the online softmax unit's architecture.
Given $\boldsymbol{x}_i,m_{i-1},l_{i-1}$, it conducts online softmax operation (i.e., Line 6-9 in Algorithm~\ref{algo:tiled-attention}) and produces $m_{i},\boldsymbol{p}_i,\alpha,l_{i}$ for subsequent attention computation.
The vector unit performs element-wise operations such as residual, normalization, and activation functions.
The reduction unit performs partial sum accumulation. As Fig.~\ref{fig:pe-arch}-(c) shows, apart from a vector scaling \& accumulation unit, it includes a factor generation unit (FGU) that produces scaling factors in the reduction version of tiled attention 
($\frac{e_1}{l},\frac{e_2}{l}$ in Eq.~(\ref{eq:fd-factor-gen})-(\ref{eq:fd-sum})).
For standard reduction, FGU's outputs will be filtered by the factor selection unit.
Each PE’s SRAM buffer is divided into a compute buffer and a transfer buffer. The compute buffer stores tensor tiles and partial sums during computation,
while the transfer buffer is used for streaming prefill KV cache into HB-Devices (details discussed later).

\begin{figure}
    \centering
    \includegraphics[width=0.99\columnwidth]{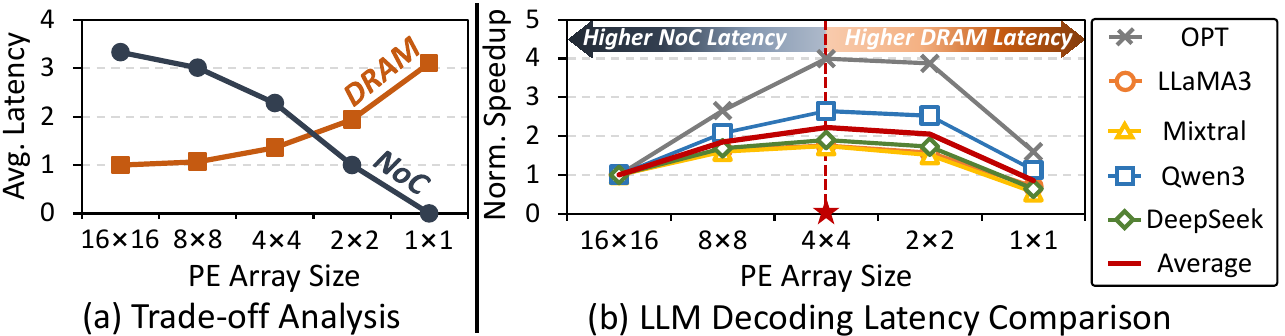}
    \vspace{-1.5em}
    \caption{PE Granularity Exploration Analysis.}
     \vspace{-0.5em}
    \label{fig:pe-exploration}
\end{figure}

\subsection{Design Considerations and Challenges}
\label{sec:design-consideration}

\noindent\textbf{Integration Overhead:}
Driving the large number of I/O pins provided by hybrid bonding requires numerous memory controllers, which encroaches logic die's area.
A recent study reports that HB controllers can occupy up to 40\% of the logic die area under 40nm for edge scenarios~\cite{li2025h2}.
This overhead decreases with the advancement of technology node. According to our industry collaborators, HB controllers only incur about 10\% area under 12nm when each bank exposes 256 I/O pins (i.e., 1024 pins aggregated from vertically aligned banks across four stacked memory dies). 
This not only ensures sufficient bandwidth provision, but also preserves ample area for compute logic customization.

\noindent\textbf{PE Granularity Exploration:}
As shown in Fig.~\ref{fig:pe-exploration}-(a), varying PE array size
(or bank partition size) introduces a trade-off between the latency overhead of NoC and DRAM:
Increasing PE number enlarges NoC's mesh size, resulting in higher 
NoC transfer overhead.
However, reducing PE number increases bank number in each PE, which hurts the effective DRAM bandwidth.
This is because, due to operator size and SRAM capacity constraints, each PE accesses nearly the same data volume across all local banks for each tensor-tile fetching.
As a result, more banks per PE reduce the data volume accessed in each bank, leading to lower bandwidth utilization.

To determine the optimal PE granularity that balances this trade-off, we evaluate five representative LLMs.
Detailed settings are summarized in Sec.~\ref{sec:evaluation-setup}.
For fair comparison, we fix total compute capacity across different array sizes.
As shown in Fig.~\ref{fig:pe-exploration}-(b), extreme array sizes \textemdash~either too large (16×16) or too small (1×1) \textemdash~result in notable performance degradation. 
We adopt the 4×4 array size with the optimal performance as the final design choice.

\noindent\textbf{Challenges in Runtime Execution:}
HB-Device features \textit{spatially distributed PEs}, which must be fully considered across the full-stack design of \TitleAbbr:
\textit{\uline{(1) For Operator Execution}}, we need not only to customize \textit{intra-PE execution strategies}, but also to design \textit{inter-PE communication flows} for diverse inter-operator transfer patterns.
\uline{\textit{(2) For System Scheduling}}, we need to incorporate \textit{spatial awareness} into fine-grained KV cache management to balance both computation and communication across PEs.
The subsequent sections will elaborate on how \TitleAbbr~addresses these remaining issues.

\section{HB-Device Operator Execution}

\subsection{Intra-PE Execution}
\label{sec:intra-pe-execution}

\begin{figure}
    \centering
    \includegraphics[width=0.99\columnwidth]{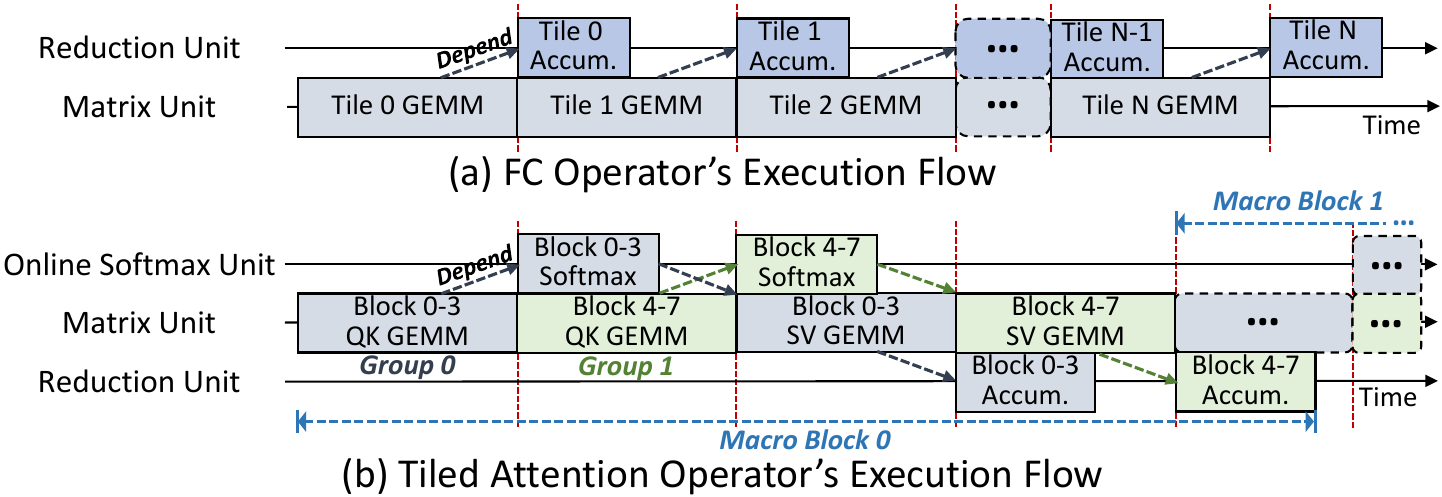}
    \vspace{-1.75em}
    \caption{Operator Execution Flow on Each PE.}
     \vspace{-1.25em}
    \label{fig:on-chip-execution}
\end{figure}

\noindent\textbf{Operator Execution:}
As shown in Fig.~\ref{fig:on-chip-execution}, \textit{\uline{for FC operators}}, after computing each tile’s GEMM results, the reduction unit performs partial-sum accumulation.
By pre-loading the next tile while computing the current tile, the partial-sum accumulation can be hidden by subsequent GEMM operation, maximizing compute resource utilization.
\textit{\uline{For tiled attention}}, to fully utilize the compute buffer, multiple blocks are merged into a larger \textit{macro block} in each iteration (e.g., 8 blocks are merged in Fig.~\ref{fig:on-chip-execution}-(b)).
Since each tiled attention iteration contains two GEMMs ($\boldsymbol{Q} \times \boldsymbol{K}_i^{\top}$ and $\boldsymbol{p}_i \times \boldsymbol{V}_i$) and two vector operations (online softmax and output accumulation), each macro block is further divided into two groups (e.g., block 0-3 and 4-7 in Fig.~\ref{fig:on-chip-execution}-(b)). 
By alternating the execution of matrix and vector operations across the two groups, vector operations' latency can be efficiently hidden.
Furthermore, by combining this intra–macro-block overlap with inter–macro-block double buffering, the explicit overhead of vector operations is negligible.
\textit{\uline{For elementwise operators}}, once the data are ready in the compute buffer, the corresponding vector processing unit can be invoked to complete the computation.

\begin{figure}
    \centering
    \includegraphics[width=0.99\columnwidth]{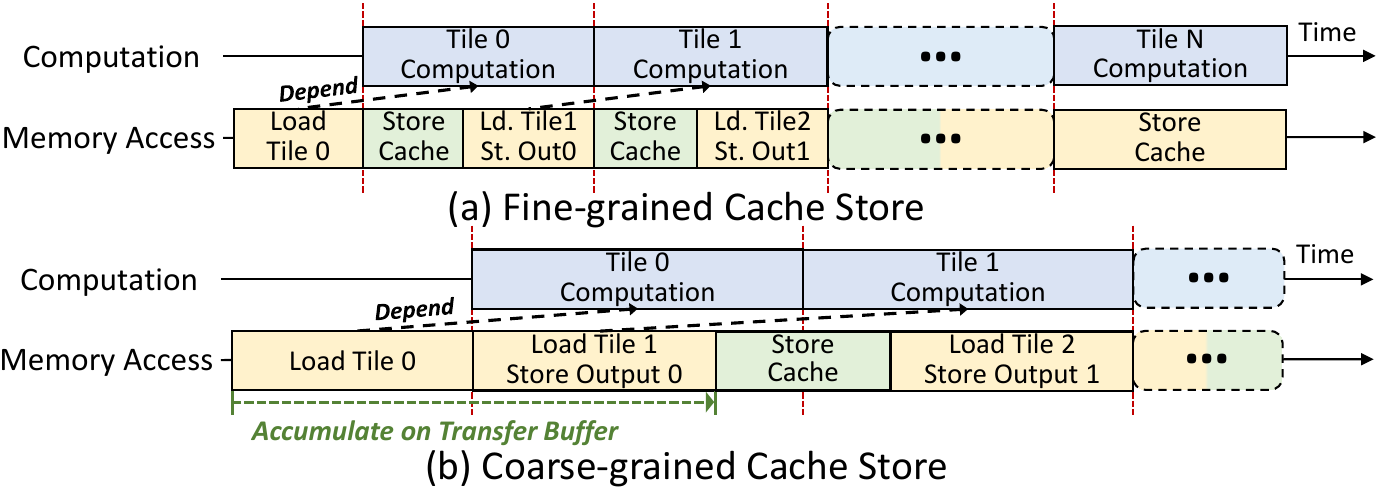}
    \vspace{-1.75em}
    \caption{Timeline of Prefill KV Cache Storing on Each PE.}
     \vspace{-0.5em}
    \label{fig:cache-transfer-overlap}
\end{figure}

\noindent\textbf{Prefill KV Cache Transfer:}
Since HB controllers are embedded within distributed PEs, streaming prompt KV cache to DRAM may interfere with the ongoing computation.
To mitigate this issue, we introduce a transfer buffer to temporarily hold incoming prompt KV cache.
Fig.~\ref{fig:cache-transfer-overlap} shows the KV cache storing timeline.
When PE's DRAM access latency is shorter than
computation latency, we exploit this time slack to perform fine-grained storing,
minimizing latency overhead.
On the other hand, When the PE is memory-bound or the KV cache arrival rate is high, we adopt the coarse-grained strategy to improve bandwidth utilization through bulk data transfers:
Incoming KV cache is first accumulated on the transfer buffer, then collectively written to the DRAM. The accumulation threshold is dynamically adjusted based on the KV cache arrival rate and the memory bandwidth utilization.
When a request arrives, the device router transfers KV cache to PEs 
designated by the policy in Sec.~\ref{sec:cache-allocation}.
Each PE dynamically selects the most suitable 
storing strategy according to the current execution status.
Once the transfer completes,
the request is added to the decoding batch via iteration-level scheduling and computed using continuous batching~\cite{yu2022orca}.

\subsection{Inter-PE Operator Partition \& Communication}
\label{sec:inter-PE}

This sub-section elaborates on \TitleAbbr's design using 4×4 mesh as an example.
Note that, as all inter-PE communication is built on \textbf{\textit{collective primitives}}, they can be generalized to \textit{\textbf{arbitrary PE array sizes and NoC topologies}}.

\begin{figure}
    \centering
    \includegraphics[width=0.99\columnwidth]{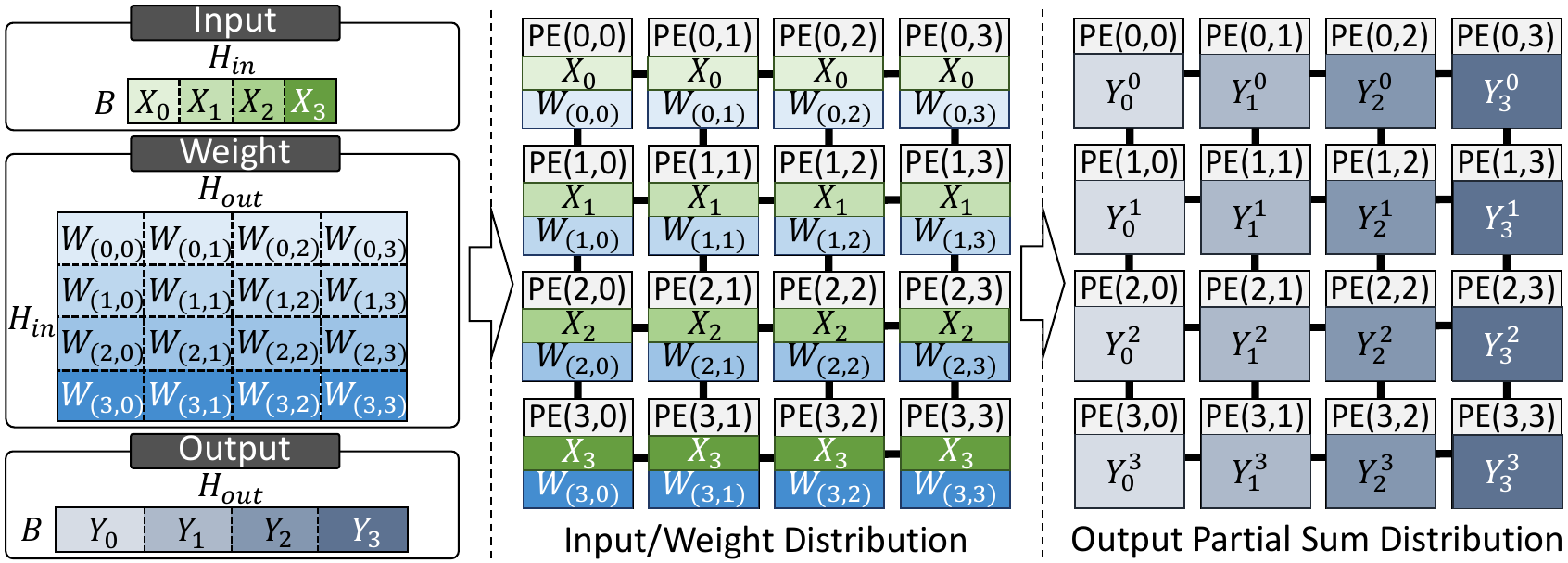}
    \vspace{-1.5em}
    \caption{Inter-PE FC Operator Partition Strategy.}
     \vspace{-0.5em}
    \label{fig:fc-partition}
\end{figure}

\noindent\textbf{Operator Partition:}
\textit{\uline{(1) For FC Operators}}, as shown in Fig.~\ref{fig:fc-partition}, given the runtime variability of decoding batch size under continuous batching~\cite{yu2022orca}, we partition only the hidden dims ($H_{in}, H_{out}$) along the PE array's two dimensions, leaving the batch size ($B$) intact.
Accordingly, the input tensor is partitioned along $H_{in}$ and replicated in accordance with weight tile mapping.
Each PE's local GEMM computation produces a partial sum of the output sub-vector 
(split along $H_{out}$), which will be reduced based on the data distribution demands of subsequent operators.
\textit{\uline{(2) For Tiled Attention}}, each request's historical KV cache is split into KV blocks (as shown in Fig.~\ref{fig:tiled-attention}-(b)) and distributed across PEs.
New KV vectors generated during decoding are appended to the last partially filled block.
All attention heads on a given device adopt the same block allocation (discussed later).
Given the block mapping, each PE first computes local attention partial sums using the iterative method in Algorithm~\ref{algo:tiled-attention}. 
These partial sums are then accumulated via Eq.~(\ref{eq:fd-factor-gen})-(\ref{eq:fd-sum}) to produce the final result.
\textit{\uline{(3) For Elementwise Operators}}, their computation can be fused into the inter-PE communication procedure.
The following paragraphs will elaborate on \TitleAbbr's inter-PE communication design for LLM's different operator blocks.

\begin{figure}
    \centering
    \includegraphics[width=0.99\columnwidth]{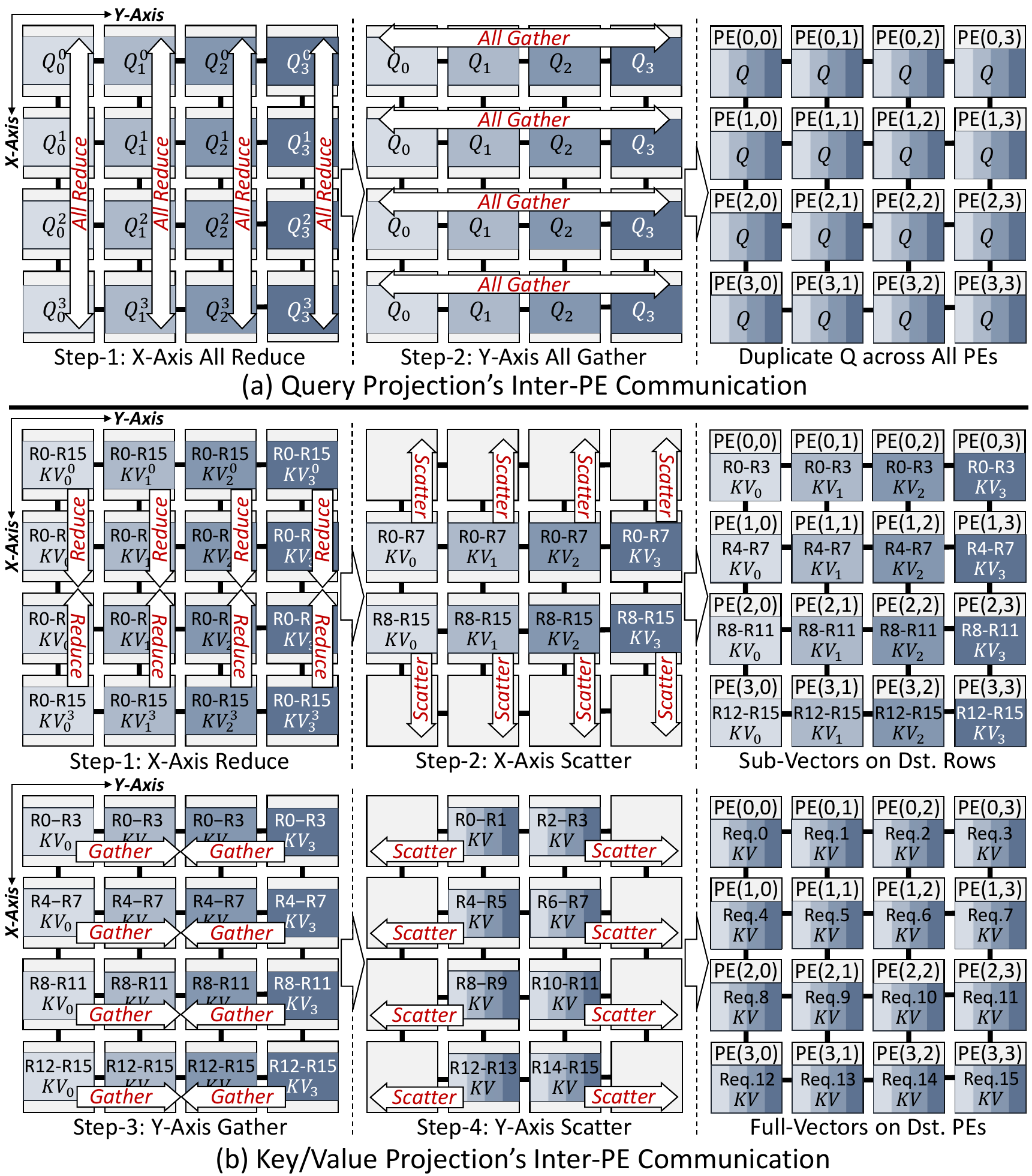}
    \vspace{-1.5em}
    \caption{Query/Key/Value Projection FC's Inter-PE Communication.}
     \vspace{-0.5em}
    \label{fig:qkv-fc-comm}
\end{figure}

\noindent\textbf{Attention Block (MHA/GQA/MQA):}
\textit{\uline{(1) Q/K/V Projection to Tiled Attention}}:
For query projection, since every PE may hold KV blocks, the complete query vector needs to be replicated across all PEs for local tiled attention.
The corresponding inter-PE communication flow is depicted in Fig.~\ref{fig:qkv-fc-comm}-(a).
Given FC's partial sum distribution, we first perform a 1D all-reduce along the X-axis to reconstruct the complete sub-vectors.
Then, through a 1D all-gather along the Y-axis, we can duplicate full query vectors across all PEs.

Unlike query vectors, key/value vectors only need to be sent to the PE holding the corresponding request's last block.
To achieve this, we employ the strategy shown in Fig.~\ref{fig:qkv-fc-comm}-(b).
In this example, we assume the batch contains 16 requests (R0-R15), with PE(x,y) storing request 4x+y's last block.
Along the X-axis, PEs in each column first send their partial sums to the central PEs, performing in-transit reduction.
The central PEs then scatter sub-vectors based on the X-axis (row) ID of the destination PE for each request.
For even-sized meshes, we can leverage the two symmetric minimum-depth trees and route requests to the root node closer to their target rows, thereby optimizing the hop count.
In Fig.~\ref{fig:qkv-fc-comm}-(b)'s example, PE(1,y) and PE(2,y) serve as the root nodes. Requests targeting rows 0–1 are reduced on PE(1,y), while those targeting rows 2–3 are reduced on PE(2,y).
In this way, Step-2's scatter can be completed in at most a single hop.
Once each row’s PEs have received corresponding sub-vectors, a similar procedure is applied along the Y-axis:
We first gather these sub-vectors on central PEs, then scatter complete vectors to target PEs,
completing the key/value vector delivery.

\begin{figure}
    \centering
    \includegraphics[width=0.99\columnwidth]{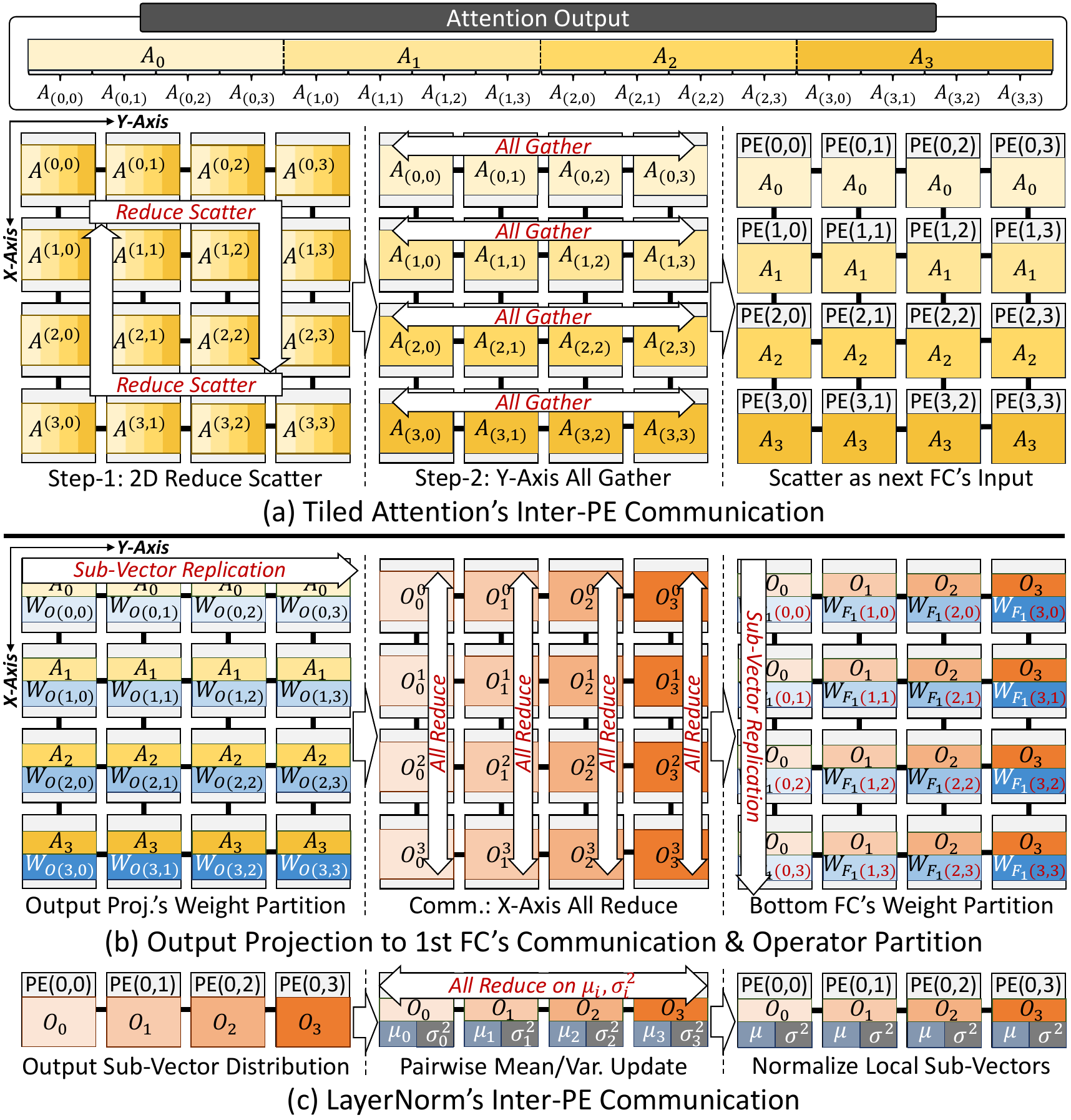}
    \vspace{-1.5em}
    \caption{Tiled Attention and Output Projection FC's Inter-PE Communication.}
     \vspace{-0.5em}
    \label{fig:attn-oproj-comm}
\end{figure}

\textit{\uline{(2) Tiled Attention to Output Projection}}:
After local tiled attention, each PE holds a partial sum of full attention output vectors (denoted as $A^{(x,y)}$).
To accumulate them and scatter the full vector according to the FC-input pattern in  Fig.~\ref{fig:fc-partition}, we employ the communication scheme shown in Fig.~\ref{fig:attn-oproj-comm}-(a).
We first perform a 2D reduce-scatter over the entire array, designating PE(x,y) as the accumulation endpoint for vector chunk $A_{(x,y)}$. After this step, we only need a single all-gather along the Y-axis to achieve the target data distribution.

\textit{\uline{(3) Output Projection to FFN/MoE's First FC}}:
As shown in Fig.~\ref{fig:attn-oproj-comm}-(b), once each PE completes its local output projection tile, a 1D all-reduce along the X-axis replicates the resulting sub-vectors across each column.
By pre-allocating the next FC's weight tile (x,y) to PE(y,x), we can directly compute it after the 1D all-reduce. 
For inter-block residual, pre-allocating the block input as $O_i$'s distribution in 
Fig.~\ref{fig:attn-oproj-comm}-(b) can achieve in-place computation.
For normalization, if the LLM adopts LayerNorm~\cite{ba2016layernormalization}, each vector's mean and variance are required.
They can be obtained via the pairwise algorithm~\cite{tile-norm}.
Formally, if two sub-vectors have means $\mu_1,\mu_2$ and variances $\sigma_1^2,\sigma_2^2$, the combined mean $\mu$ and variance $\sigma^2$ are:
\begin{equation}
\label{eq:pairwise-norm}
    \textstyle{\mu=\frac{n_1\mu_1+n_2\mu_2}{n_1+n_2},
    \sigma^2=\frac{n_1\sigma_1^2+n_2\sigma_2^2}{n_1+n_2}+\frac{n_1n_2(\mu_1-\mu_2)^2}{(n_1+n_2)^2}}
\end{equation}
where $n_1,n_2$ are element counts of the two sub-vectors.
Hence, as shown in 
Fig.~\ref{fig:attn-oproj-comm}-(c), performing a 1D all-reduce on each sub-vector's mean and variance can yield the full vector's statistics for LayerNorm.
For RMSNorm~\cite{zhang2019rootmeansquarelayer}, since it only requires full vector's L2 norm, an all-reduce over each sub-vector's squared sum is sufficient for this operator.

\noindent\textbf{FFN \& MoE Block:}
Since the FFN block consists of chained FC operators, by adopting the same weight tile placement as Fig.~\ref{fig:attn-oproj-comm}-(b), we only need a single 1D all-reduce to achieve the required data layout.
The activation function can be applied in place after the all-reduce.
For GLU block, by applying identical partition to the two bottom FCs, corresponding elements reside on the same PE after the 1D all-reduce, enabling in-place elementwise multiplication.
For MoE block, expert selection is runtime-dependent. To maximize resource utilization, we distribute each expert across all PEs using the same partition scheme as the FFN block.
During execution, the global controller adopts MoE router's output to inform PEs which expert kernels to compute. 
The selected experts are then processed in the same manner as the FFN.

\begin{figure}
    \centering
    \includegraphics[width=0.99\columnwidth]{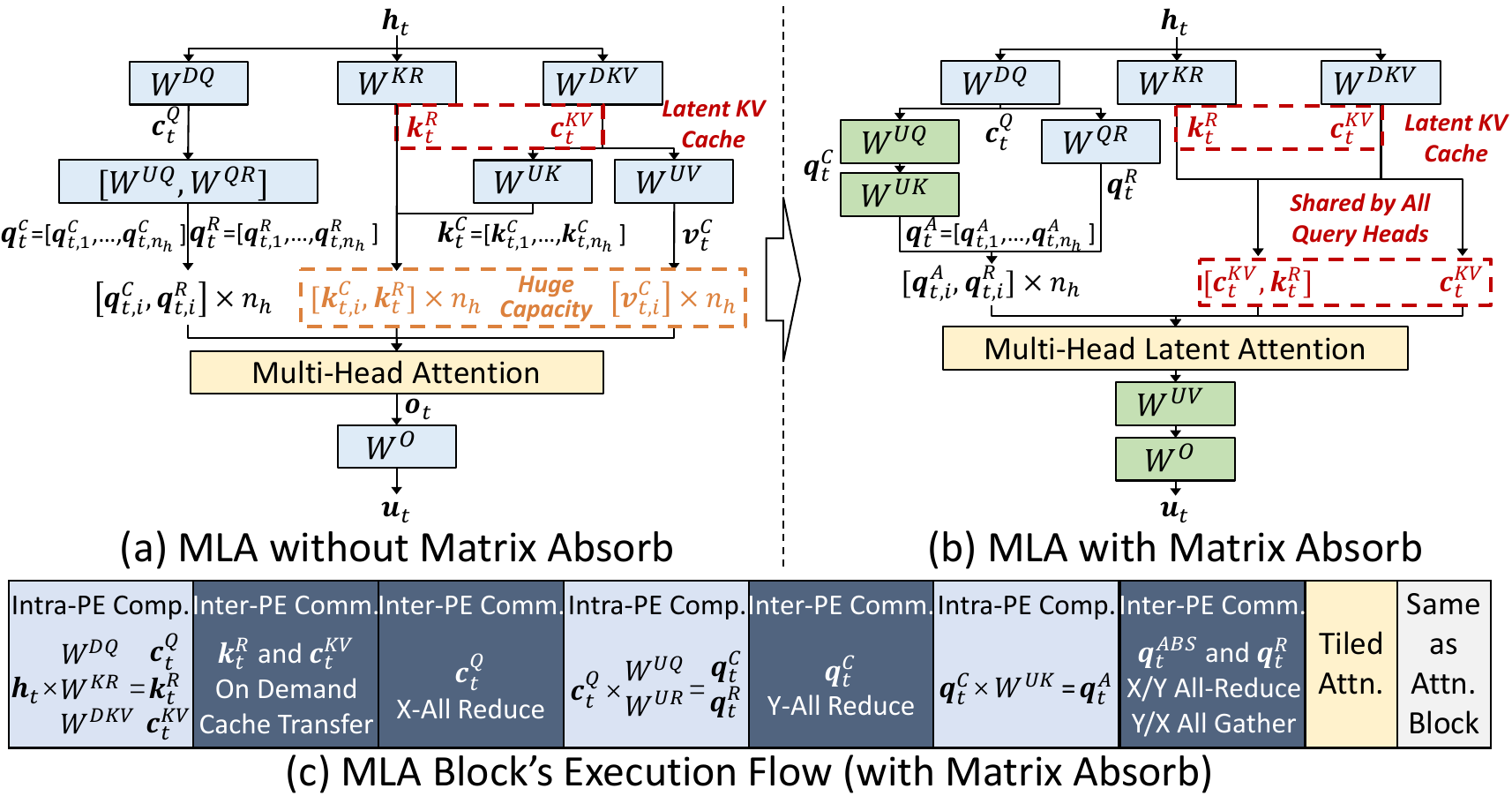}
    \vspace{-1.5em}
    \caption{MLA Block's Inter-PE Communication.}
     \vspace{-0.5em}
    \label{fig:mla-comm}
\end{figure}

\noindent\textbf{MLA Block:}
Unlike the attention block, MLA block introduces additional FC operators to generate latent KV cache. As shown in Fig.~\ref{fig:mla-comm}-(a), the input $\boldsymbol{h}_t$ is first projected through three FCs, where $\boldsymbol{k}_t^{R},\boldsymbol{c}_t^{kv}$ are persisted as the latent KV cache. 
In naïve MLA, these latent vectors are first reconstructed to all heads' full-resolution query/key/value vectors, followed by standard MHA computation.
This approach generates huge volume of full KV cache as intermediate results, undermining MLA's benefits.
Matrix absorption~\cite{liu2024deepseek} is proposed to address this issue. As shown in Fig.~\ref{fig:mla-comm}-(b), the reconstruction matrices $W^{UK},W^{UV}$ can be algebraically absorbed into $W^{UQ},W^{O}$ via GEMM's associative property.
Without loss of generality, we demonstrate this procedure using a single token $t$ with $n_h=1$. Assuming this token's softmax score is $s_t$, we have:
\begin{equation}
\begin{aligned}
\boldsymbol{q}_t^{C}(\boldsymbol{k}_t^{C})^{\top}
&=(\boldsymbol{c}_t^{Q}W^{UQ})(\boldsymbol{c}_t^{KV}W^{UK})^{\top}\\
&=(\boldsymbol{c}_t^{Q}W^{UQ})((W^{UK})^{\top}(\boldsymbol{c}_t^{KV})^{\top})\\
&=\textcolor[rgb]{0.329, 0.510, 0.208}{((\boldsymbol{c}_t^{Q}W^{UQ})(W^{UK})^{\top})}(\boldsymbol{c}_t^{KV})^{\top}
\end{aligned}
\end{equation}
\begin{equation}
\begin{aligned}
\boldsymbol{u}_t=\boldsymbol{o}_tW^O
=(s_t\cdot\boldsymbol{v}_t^C)W^O
&=(s_t\cdot(\boldsymbol{c}_t^{KV}W^{UV}))W^O\\
&=((s_t\cdot\boldsymbol{c}_t^{KV})\textcolor[rgb]{0.329, 0.510, 0.208}{W^{UV}})\textcolor[rgb]{0.329, 0.510, 0.208}{W^O}
\end{aligned}
\end{equation}
In this way, all query heads can operate on the shared latent KV cache, without explicitly reconstructing full KV tensors.

MLA block's execution flow is shown in Fig.~\ref{fig:mla-comm}-(c).
All PEs first compute $W^{DQ},W^{KR},W^{DKV}$ following Fig.~\ref{fig:fc-partition}. 
Then, latent KV cache $\boldsymbol{k}_t^R,\boldsymbol{c}_t^{KV}$ are transferred on demand via Fig.~\ref{fig:qkv-fc-comm}-(b), while $\boldsymbol{c}_t^Q$ undergoes X-axis all-reduce (Fig.~\ref{fig:attn-oproj-comm}-(b)) for subsequent FCs.
After computing $W^{UQ},W^{QR}$, $\boldsymbol{q}_t^C$ undergoes Y-axis all-reduce (Fig.~\ref{fig:attn-oproj-comm}-(b)) and is multiplied with $W^{UK}$ to produce $\boldsymbol{q}_t^A$.
Then, $\boldsymbol{q}_t^A,\boldsymbol{q}_t^R$ are fully replicated across all PEs following Fig.~\ref{fig:qkv-fc-comm}-(a).
Finally, PEs perform distribtued tiled attention (Fig.~\ref{fig:attn-oproj-comm}-(a)) and chained FC ($W^{UV}$,$W^{O}$) (Fig.~\ref{fig:attn-oproj-comm}-(b)) to complete MLA block's computation.

\noindent\textbf{Communication Implementation:}
The whole procedure involves two categories of collective primitives: (1) For all-reduce/all-gather/reduce-scatter, we adopt the TidalMesh algorithm~\cite{tidalmesh}, which achieves state-of-the-art performance on mesh topology compared with ring-/tree-based methods. (2) For reduce/gather/scatter, we employ the minimum-spanning-tree-based algorithm~\cite{cctheory} to fully utilize all available links.

\section{\TitleAbbr~System Design}

\subsection{Spatially-Aware KV Cache Allocation}
\label{sec:cache-allocation}

\noindent\textbf{Demand Analysis:}
HB-Device contains two load balance demands:
\textit{\uline{(1) Compute/Storage load}}: 
It enforces a balanced distribution of attention workloads across all PEs on the HB-Device.
For each PE, as discussed in Sec.~\ref{sec:intra-pe-execution}, online softmax and partial sum accumulation can be efficiently overlapped by GEMM operations.
Therefore, tiled attention's compute load can be estimated by the corresponding matrix multiplication workloads.
Since they are proportional to token number~\cite{zhong2024distserve},
\textbf{\textit{we can use token number as the metric}} to guide compute load balance, which simultaneously serves as a proxy for KV cache storage load balance.
\textit{\uline{(2) Inter-PE communication load}}: 
The cost of key/value vector transmission depends on last-block distribution of ongoing requests.
As shown in Fig.~\ref{fig:qkv-fc-comm}-(b), \textbf{\textit{placing last blocks on central PEs}} used by X-Reduce/Y-Gather (PE(1,1)/(1,2)/(2,1)/(2,2)) eliminates X/Y-axis scatter in Step-2/4.
Additionally, \textbf{\textit{placing last blocks symmetrically}} (e.g., (1,1)-(2,1) for X-axis, (1,1)-(1,2) for Y-axis) can further balance traffic loads across parallel NoC links.

\begin{algorithm}[t]
\setstretch{1.0}
\footnotesize
\caption{\small Spatially-Aware KV Block Allocation}
\label{algo:block-allocation}

\SetKwIF{If}{ElseIf}{Else}{if}{,}{else if}{else}{end}

\SetKwFunction{FMain}{Allocate}
\SetKwProg{Fn}{Function}{}{}

\Fn{\FMain{$Req$, $b$, $PEState$}  \textcolor{blue}{\textnormal{// Main allocator}}}{

\For(\hfill \textcolor{blue}{// {Allocate full blocks}}){$i=0$ to $N_{full}-1$}{

$State_{F} \leftarrow Sort(PEState, \text{\texttt{CMPFull}})$

$PEState.UpdateLoad(State_F[0],Req.ID,b)$

}

$State_{L} \leftarrow Sort(PEState, \text{\texttt{CMPLast}})$  \textcolor{blue}{// {Allocate last block}}

$PEState.UpdateLoad(State_L[0],Req.ID,T_{last})$

}

\SetKwFunction{FMain}{CMPFull}
\SetKwProg{Fn}{Function}{}{}
\Fn{\FMain{$PE_1$, $PE_2$}  \textcolor{blue}{\textnormal{// Full block comparator}}}{
    \lIf{$PE_1.T_{sum} \ne PE_2.T_{sum}$}{
        \Return $PE_1.T_{sum} < PE_2.T_{sum}$
    }
    \Return $PE_1.D_{L1} > PE_2.D_{L1}$
}


\SetKwFunction{FMain}{CMPLast}
\SetKwProg{Fn}{Function}{}{}
\Fn{\FMain{$PE_1$, $PE_2$}   \textcolor{blue}{\textnormal{// Last block comparator}}}{
    \lIf{$PE_1.T_{sum} \ne PE_2.T_{sum}$}{
        \Return $PE_1.T_{sum} < PE_2.T_{sum}$
    }
    \lIf{$PE_1.D_{L1} \ne PE_2.D_{L1}$}{
        \Return $PE_1.D_{L1} < PE_2.D_{L1}$
    }
    \lIf{$PE_1.N_{last} \ne PE_2.N_{last}$}{
        \Return $PE_1.N_{last} < PE_2.N_{last}$
    }
    \lIf{$PE_1.L_{mst}^X \ne PE_2.L_{mst}^X$}{
        \Return $PE_1.L_{mst}^X < PE_2.L_{mst}^X$
    }
    \Return $PE_1.L_{mst}^Y < PE_2.L_{mst}^Y$
}

\end{algorithm}

\noindent\textbf{Allocation Strategy:}
To satisfy these goals, we propose the strategy listed in Algorithm~\ref{algo:block-allocation}. 
The main entry \texttt{allocate} takes three inputs: request information $Req$, KV block size $b$, and all PEs' load states $PEState$.
It has three steps:

\uline{\textit{Step-1}}:
It derives full block number $N_{full}$ and the last (partially filled) block's token count $T_{last}$ from total token number ($Req.T$) and KV block size $b$ (\textbf{Line 2}).
$T$ is set to the prompt length for newly arrived requests and to 1 for ongoing requests requiring new blocks.

\uline{\textit{Step-2}}:
For each full block, it first sorts PEs based on $PEState$ and selects the top-ranked PE to store the block (\textbf{Line 4}).
$PEState$ is updated accordingly to guide subsequent decisions (\textbf{Line 5}).
The ranking rule \texttt{CMPFull} is defined as: (1) PEs with fewer tokens $T_{sum}$ have higher priority to balance the compute/storage load (\textbf{Line 9}). (2) If two PEs' token counts are equal, the PE with larger 
Manhattan (L1) distance $D_{L1}$ to its nearest central PE is preferred (\textbf{Line 10}).
In the example of Fig.~\ref{fig:qkv-fc-comm}-(b), the closest central PE to PE(0,0) is PE(1,1), yielding $D_{L1}$ to $|0-1|+|0-1|=2$.
This rule reserves PEs closer to the center for unfilled blocks, reducing the inter-PE transfer overhead in Fig.~\ref{fig:qkv-fc-comm}-(b).

\uline{\textit{Step-3}}:
It adopts similar sort-and-update procedure to the last block under ranking criterion \texttt{CMPLast} (\textbf{Line 6-7}):
(1) PEs with fewer tokens $T_{sum}$ are prioritized to balance compute/storage load (\textbf{Line 12}).
(2) If $T_{sum}$ tie, PEs with smaller $D_{L1}$ are ranked higher to reduce KV vector's transfer cost (\textbf{Line 13}).
(3) If both metrics are equal, PEs with fewer unfilled blocks $N_{last}$ are preferred to balance intra-block fragmentation among PEs (\textbf{Line 14}).
(4) Finally, PEs are ranked by the data volume along their X/Y-axis transfer paths $L_{mst}^{X}/L_{mst}^{Y}$, so as to balance the transfer cost across parallel paths of spanning trees in 1D mesh (\textbf{Line 15-16}).

\begin{figure}
    \centering
    \includegraphics[width=0.99\columnwidth]{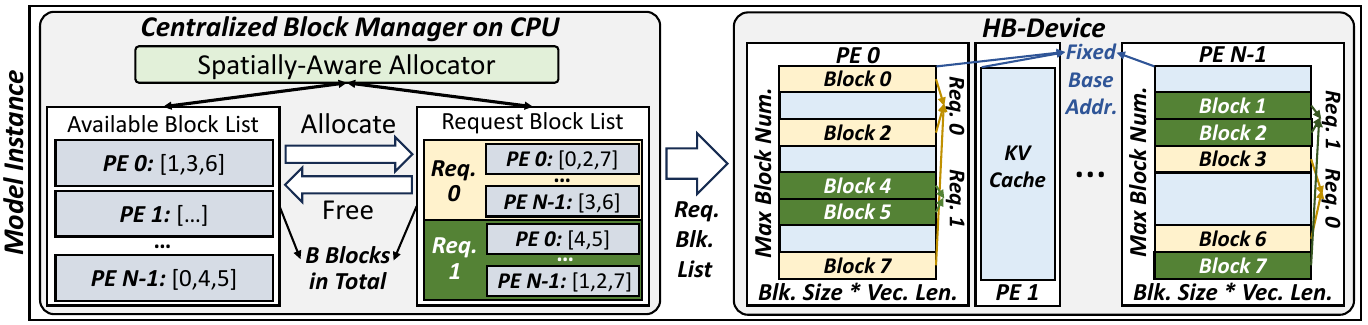}
    \vspace{-1.5em}
    \caption{\TitleAbbr's Block Manager Implementation.}
     \vspace{-0.5em}
    \label{fig:block-allocator}
\end{figure}

\noindent\textbf{Block Manager Implementation:}
As shown in Fig.~\ref{fig:block-allocator}, each PE pre-allocates the KV cache tensor during initialization according to serving requirements.
With a maximum of $M$ blocks per PE, block size $b$, and KV vector length $H$, the tensor has shape ($M,bH$).
Unlike prior NMP designs that reserve full-context-window KV cache for each request,
\TitleAbbr~allocates blocks based on the current request length and distributes them across all PEs using Algorithm~\ref{algo:block-allocation}.
Since each KV cache tensor's base address is fixed, the address of slot j in block i can be computed as: $base+(i \cdot b + j) \cdot H$.
Therefore, during inference, it suffices to provide each request's block ID list to resolve KV blocks' addresses for tiled attention.

For each model instance, block tables are centrally maintained on the CPU.
The \textit{available block list} tracks free block IDs on each PE, while the \textit{request block list} records the block IDs assigned to each request on each PE.
When allocating a new block, the spatially-aware allocator (Algorithm~\ref{algo:block-allocation}) moves a block ID from the available list to the corresponding request list. After decoding completes, all block IDs associated with the request are released back to the available list.
Since each block ID exists exclusively in either the available or request list, the space overhead of the block manager scales linearly with the total block number $MN$ ($N$ is PE number), consistent with implementations in current GPU systems~\cite{sglang,kwon2023efficient}.

\noindent\textbf{Scheduling Overhead Analysis:}
To select the optimal PE for each block, PE states can be maintained in priority queues  under \texttt{CMPFull}/\texttt{CMPLast} comparison rules.
Given $\mathit{P}$ PEs, the time complexity of priority-queue-based PE selection is $\mathit{O}(\log \mathit{P})$~\cite{priority-queue}.
Considering practical architectures contain only a small number of PEs (e.g., 16 in our final design), the runtime overhead of Algorithm~\ref{algo:block-allocation} is negligible, comparable to that of the SGLang block manager in our evaluation~\cite{sglang}.

\subsection{Multi-Device Scaling}

\TitleAbbr's full-stack design is compatible with
multi-device model parallelism:
\textit{\uline{(1) Operator Execution}}:
For TP/PP, operators are first split across devices based on TP/PP sizes.
During computation, inter-device collective communication is performed via the device router.
Once receiving the next operator's input, each device scatters it via the router NoC according to the partition in Sec.~\ref{sec:inter-PE}, then continue operator execution.
For EP, each device holds a subset of non-partitioned experts. Thus, after receiving inputs, each device can directly compute selected experts as discussed in Sec.~\ref{sec:inter-PE}. 
This mechanism supports any expert-mapping and input-routing policy.
\textit{\uline{(2) KV Cache Management}}: 
Since  each device retains an equal share of KV heads after model partition, we can adopt a centralized block manager to uniformly allocate KV blocks via Algorithm~\ref{algo:block-allocation}, similar to vLLM's practice~\cite{kwon2023efficient}.

\section{Evaluation}
\label{sec:evaluation}

\begin{table}[t]
\centering

\caption{Model Configurations Used for Evaluation}
\label{tab:model-config}
\vspace{-0.5em}
\resizebox{0.475\textwidth}{!}{

\begin{tabular}{|c|c|c|c|c|}
\hline

Model & Layer & (Hidden, Interm.) & (Q head, KV head)  & Expert \\ \hline

OPT 66B & 64 & (9216, 36864) & (72, 72) & -- (Dense) \\ \hline

LLaMA3 70B & 80 & (8192, 28672) & (64, 8) & -- (Dense) \\ \hline

Mixtral 8×22B & 56 & (6144, 16384) & (48, 8) & 8 experts, top-2 \\ \hline

Qwen3 30B-A3B & 48 & (2048, 768) & (32, 4) & 128 experts, top-8 \\ \hline

DeepSeek 236B & 60 & (5120, 1536) & (128, 128-full/1-latent) & 160 experts, top-8 \\ \hline

\end{tabular}

\vspace{-1em}
}
\end{table}

\subsection{Evaluation Methodology}
\label{sec:evaluation-setup}

\noindent\textbf{Benchmarks:}
As listed in Table~\ref{tab:model-config}, we adopt
OPT 66B~\cite{zhang2022opt}, LLaMA3 70B~\cite{dubey2024llama}, Mixtral 8×22B~\cite{jiang2024mixtral}, Qwen3 30B-A3B~\cite{yang2025qwen3}, and DeepSeek 236B~\cite{liu2024deepseek} for evaluation, which cover both dense and MoE models with varying attention arithmetic intensities. 
All models use FP16 data type.
For decoding latency comparison, due to capacity constraints, we set max batch size and context length to (32, 4K) for OPT/DeepSeek models, and to (64, 16K) for the other models.
For serving performance comparison, following prior works~\cite{kwon2023efficient,zhong2024distserve}, we sample request lengths from the production trace released by MoonCake~\cite{qin2025mooncake} and generate arrival timestamps using Poisson process under varying request arrival rates.

\noindent\textbf{Baselines:}
We compare \TitleAbbr~with four baselines: \textbf{\uline{(1)}} A100-80GB GPUs connected with 600GB/s NVLink (GPU). We use SGLang~\cite{sglang} with FlashInfer~\cite{ye2025flashinfer} backend for evaluation.
\textbf{\uline{(2)}} In-die NMP devices (In-die-NMP). We replace A100's HBM cubes to AttAcc's bank-PIM design~\cite{park2024attacc}, which integrates a 16-MAC-FPU @ 666MHz near each bank while preserving the cube's full capacity.
\textbf{\uline{(3)}} Duplex-based devices~\cite{yun2024duplex} (Duplex). 
We replace A100’s HBM cubes to Duplex cubes, which expose additional TSVs to offer 4× NMP bandwidth.
Besides, it places 32 512-MAC-FPUs @ 650MHz on the cube's logic die, achieving higher compute-bandwidth ratio (8:1) than all existing in-die NMP without sacrificing capacity.
\textbf{\uline{(4)}} HB-Devices with non-tiled attention (HB-non-Tiled).
To assess the benefits of our tiled attention design, we evaluate HB-Device's performance by adopting the execution flow of Stratum~\cite{pan2025stratum}.
Since Stratum uses the same number of PEs as HB-Device, and its ring topology is a subset of mesh topology, we can directly implement its operator execution dataflow on our HB-Device.

All instances are set to TP=8(+EP=8 for MoE models) in eight-device nodes. 
In serving comparison, GPU/In-die-NMP/Duplex are evaluated under both mixed-batching and disaggregated clusters, while HB-non-tiled/\TitleAbbr~are tested under disaggregated cluster.
\TitleAbbr's block size is 64.
Since NMP offers no prefill speedup, we adopt A100 GPUs to all prefill instances in disaggregated clusters.
For fair comparison, we set each node's device interconnect to 600GB/s via NVLink fusion~\cite{nvlink}, and inter-node link to 8×200Gbps NICs.

\noindent\textbf{Simulation:}
For decoding execution, we extend Ramulator2~\cite{luo2023ramulator} and BookSim~\cite{jiang2013detailed} to simulate HB-Devices.
For NMP baselines, we refer to AttAcc and Duplex simulators~\cite{attacc-sim,duplex-sim}, and inject GPU operator performance profiled on real-machine for end-to-end simulation. GPU's energy is profiled via PyNVML~\cite{pyNVML}. 
NVLink's energy is 1.3pJ/bit according to its technical report~\cite{nvlink-c2c}.
For serving simulation, we extend DistServe's simulator~\cite{simdistserve} to support both mixed and disaggregated clusters. We inject the simulated/profiled inference latency to evaluate different designs' serving performance.

\begin{figure}
    \centering
    \includegraphics[width=0.99\columnwidth]{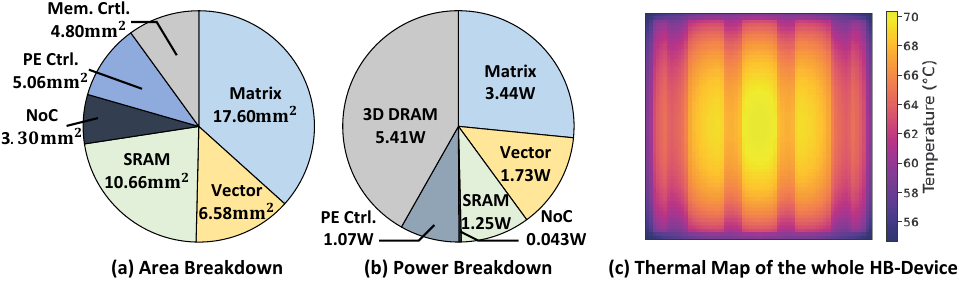}
    \vspace{-1.5em}
    \caption{Area, Power, and Thermal Analysis of HB-Device}.
     \vspace{-0.5em}
    \label{fig:ppa-breakdown}
\end{figure}

\subsection{Area, Power, and Thermal Analysis}
\label{sec:ppa}

Each HB-Device integrates four DRAM dies, each comprising 256 banks of 80MB (20GB/die).
Each bank exposes 256 I/O pins running at 0.5Gbps.
The logic die is synthesized with 12nm technology, operating at 1GHz.
The PE array is organized as 4×4 mesh according to Sec.~\ref{sec:design-consideration}.
As shown in Fig.~\ref{fig:ppa-breakdown}-(a), each PE's area is 48mm$^2$,
with 10\% consumed by the HB controller and I/O.
Matrix unit occupies 17.6mm$^2$, containing 512 16-MAC FPUs.
Online softmax/vector/reduction units occupy 6.58mm$^2$ in total.
Each PE's compute/transfer buffer sizes are both 2.5MB, with 10.66mm$^2$ area in total according to TSMC SRAM compiler.
Each NoC's area is 1.65mm$^2$, with 0.91mm$^2$ for buffers, 0.69mm$^2$ for wires,
and 0.05mm$^2$ for router.
HB-Device’s compute, memory access, and SRAM access energy are 0.43pJ/FLOP, 0.66pJ/bit~\cite{wang2023135}, and 0.019pJ/bit, respectively.

Higher temperatures degrade DRAM retention, increasing refresh frequency and reducing effective bandwidth.
To ensure reliable operation, 3D-DRAM-based accelerators must avoid excessive temperatures (typically $\le$85\textcelsius{}~\cite{yue2024exploiting,yue20253d}).
Given the per-PE power breakdown at maximum utilization in Fig.~\ref{fig:ppa-breakdown}-(b), we perform thermal analysis on
HB-Device using Hotspot-7.0~\cite{han20222}.
For 3D-integration, the vertical heat resistance of hybrid bonding interface and TSV are 1.625W/(m$\cdot$K)~\cite{oprins20203d} and 190W/(m$\cdot$K)~\cite{ren2020thermal}, respectively. 
For heat sink, we adopt advanced liquid cooling with \textasciitilde0.01W/K heat convection resistance, similar to Stratum's settings~\cite{pan2025stratum}.
As shown in Fig.~\ref{fig:ppa-breakdown}-(c), the peak temperature of HB-Device remains below 75\textcelsius{}, safely satisfying the constraint and avoiding thermal throttling.

\begin{figure*}
    \centering
    \includegraphics[width=0.99\linewidth]{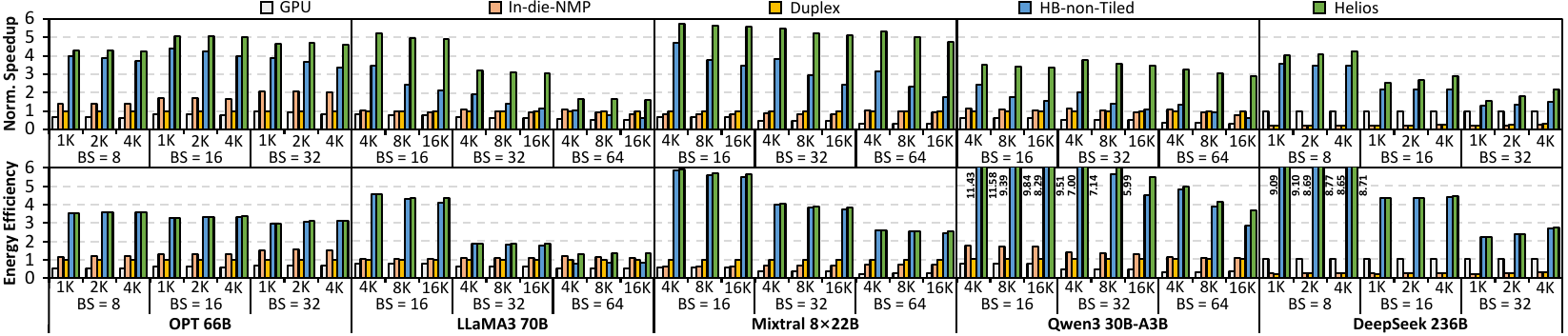}
    \vspace{-0.5em}
    \caption{Decoding Performance Comparison under Varying Request Lengths (Top: Speedup. Bottom: Energy Efficiency).}
    \vspace{-1.25em}
    \label{fig:decoding-comp}
\end{figure*}

\begin{figure*}
    \centering
    \includegraphics[width=0.99\linewidth]{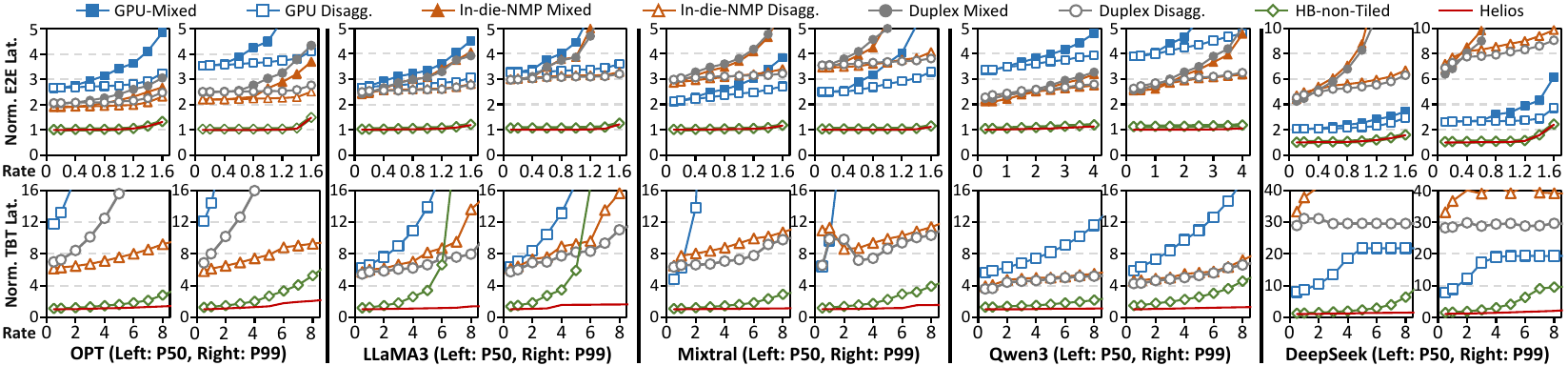} 
    \vspace{-0.5em}
    \caption{Serving Performance Comparison. (Top: E2E Latency under Full Serving. Bottom: TBT Latency under Decoding Stress Test.)}
    \vspace{-1.75em}
    \label{fig:serving-comp}
\end{figure*}

\subsection{Decoding Performance Comparison}
\label{sec:decoding-performance-comparison}

We first compare \TitleAbbr's decoding latency with all baselines. To better reflect real-world serving workloads, we sample request lengths from MoonCake's conversation dataset~\cite{qin2025mooncake} across different length ranges.
For In-die-NMP/Duplex (NMP baselines), we adopt Duplex's execution flow~\cite{yun2024duplex}, which offloads decoding attention to NMP and overlaps MoE experts between GPU and NMP.
To eliminate the impact of imbalanced expert routing on hardware evaluation, we follow Duplex~\cite{yun2024duplex} in using uniform distribution to route experts.

The results are shown in 
Fig.~\ref{fig:decoding-comp}, where we normalize DeepSeek's results to GPU and other models to Duplex.
By increasing bandwidth, NMP baselines can accelerate both MHA/GQA and MoE expert FCs (arithmetic intensity reduced by expert routing), thus outperforming GPUs on non-DeepSeek models.
Besides, in-die NMP outperfoms Duplex on MHA-based OPT due to its higher NMP bandwidth, while Duplex performs better on LLaMA3/Mixtral/Qwen3 due to its higher compute power relative to in-die NMP.
However, DeepSeek's MLA significantly boosts arithmetic intensity by sharing one latent KV head across all query heads. 
Given the limited compute power,
in-die NMP/Duplex only yield 21.6-34.5\% of GPU performance on DeepSeek.

On the other hand, by improving compute power on the holistic logic die and leveraging hybrid bonding's high bandwidth, HB-non-Tiled outperforms GPU/NMP baselines in most cases.
However,statically partitioning key/value matrices fails to accommodate the variable-length requests in real LLM serving workloads.
As the context window expands, the divergence in request lengths becomes more pronounced.
Consequently, HB-non-Tiled's performance progressively declines and even drops below NMP baselines (e.g., LLaMA3/Qwen with 16K context).
By customizing efficient execution flow for tiled attention, \TitleAbbr~fully exploits HB-Device's capabilities,
achieving 1.67/3.25× speedup (geomean) and 1.06/3.36× better energy efficiency (geomean) compared with HB-non-tiled and best-performing GPU/NMP baselines, respectively.

\subsection{Serving Performance Comparison}

\noindent\textbf{End-to-end Serving:}
We first compare with all baselines under both mixed and disaggregated clusters.
To minimize the impact of scheduler-induced load imbalance on hardware evaluation, each disaggregated cluster hosts one prefill and one decoding instance, while each mixed cluster runs two SGLang-scheduled instances.
Due to capacity limits, we truncate MoonCake's conversation dataset with prompt length $\le$8K and decoding length $\le$256.
To capture the performance impact of cluster architecture, we compare the end-to-end (E2E, both prefill and decoding) latency distribution, and plot P50/P99 trends under varying request rates in the top row of Fig.~\ref{fig:serving-comp}.

For GPU/NMP baselines, mixed clusters' latency increases sharply even under moderate pressure, exceeding that of disaggregated clusters.
Although Duplex overlaps prefill/decoding attention between GPU and NMP to alleviate mixed-batch latency, it still falls short of the benefits achieved by disaggregation.
In disaggregated clusters, decoding instances operate with small batches (average $\le$4) due to prefill throughput limitation.
In this low-load condition, in-die NMP and Duplex show comparable performance.
Small batches also exacerbate the load imbalance caused by coarse-grained KV cache management.
Although NMP baselines achieve better performance than GPU on OPT/LLaMA3/Qwen, this imbalance prevents them from outperforming GPU on Mixtral.
Besides, their limited compute power leads to the performance well below GPU on DeepSeek.
Compared with best-performing GPU/NMP baselines, \TitleAbbr~achieves only 39.5-56.9\%/32.4-65.1\% of P50/P99 E2E latency.
For HB-non-Tiled, since attention contributes marginal overhead at low load, it reports up to 6.7\%/13.7\% higher P50/P99 E2E latency than \TitleAbbr.

\noindent\textbf{Decoding Stress Test:}
To thoroughly evaluate decoding serving performance, we conduct stress tests on one decoding instance, assuming prefill throughput is sufficient enough. 
We also expand sampling range to prompt lengths $\le$16K and decoding lengths $\le$512.
The bottom row of Fig.~\ref{fig:serving-comp} shows P50/P99 distributions of time-between-tokens (TBT).
We can find that TBT performance closely aligns with decoding benchmark results.
Among GPU/NMP baselines, in-die NMP performs best on OPT, GPU on DeepSeek, and Duplex on the other models.
For low-load Mixtral scenarios (arrival rate $\le$1), NMP baselines exhibit higher TBT than GPU due to the aforementioned attention load imbalance.
Compared with best-performing GPU/NMP baselines, \TitleAbbr~achieves only 5.6-28.0\%/6.9-23.9\% of P50/P99 TBT.
For HB-non-Tiled, the inefficiency brought by non-tiled attention execution
leads to significant performance degradation under high pressure. On LLaMA, severe request accumulation pushes its performance below NMP baselines. On other models, it incurs 1.08-6.87×/ 1.17-4.86× higher P50/P99 TBT than \TitleAbbr.

\begin{figure}
    \centering
    \includegraphics[width=0.99\columnwidth]{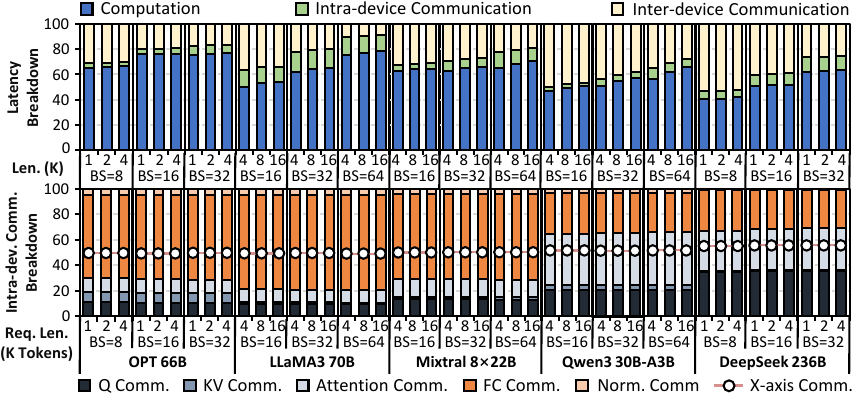}
    \vspace{-1.5em}
    \caption{Decoding Latency and Intra-device Communication Breakdown.}
     \vspace{-0.5em}
    \label{fig:communication-breakdown}
\end{figure}

\subsection{Performance Analysis}
\label{sec:perf-analysis}

\noindent\textbf{Latency Breakdown:}
To evaluate the effectiveness of our PE granularity exploration and inter-PE communication design,
we break down HB-Device's decoding latency in Sec.~\ref{sec:decoding-performance-comparison}, including PE computation, intra-device (i.e., inter-PE) communication, and inter-device communication.
As shown in the first row of Fig.~\ref{fig:communication-breakdown}, inter-PE communication accounts for only 2.94\%-16.10\% (average 7.08\%) of the total latency, validating the efficiency of our communication design.
After optimizing intra-device computation and communication overhead, NVLink-based inter-device communication accounts for 8.98\%-53.48\% (average 27.73\%).
However, as HB-Device can leverage GPU-compatible inter-device interconnect technologies, this overhead aligns with that of GPU systems and is orthogonal to our architectural design.

We further break down inter-PE communication latency. 
The bar charts show its decomposition into five communication types in Fig.~\ref{fig:qkv-fc-comm}-\ref{fig:attn-oproj-comm}, and the line plot shows the fraction of traffic along the X-axis.
Query vector replication contributes 9.63\%-35.53\% (average 15.71\%) of inter-PE communication latency, translating to only 0.35\%-4.07\% (average 1.11\%) of decoding latency, indicating that it is not a performance bottleneck.
Besides, X-axis traffic accounts for 49.21\%-55.54\% of total overhead, demonstrating that our design effectively utilizes links across different mesh dimensions.

\begin{figure}
    \centering
    \includegraphics[width=0.99\linewidth]{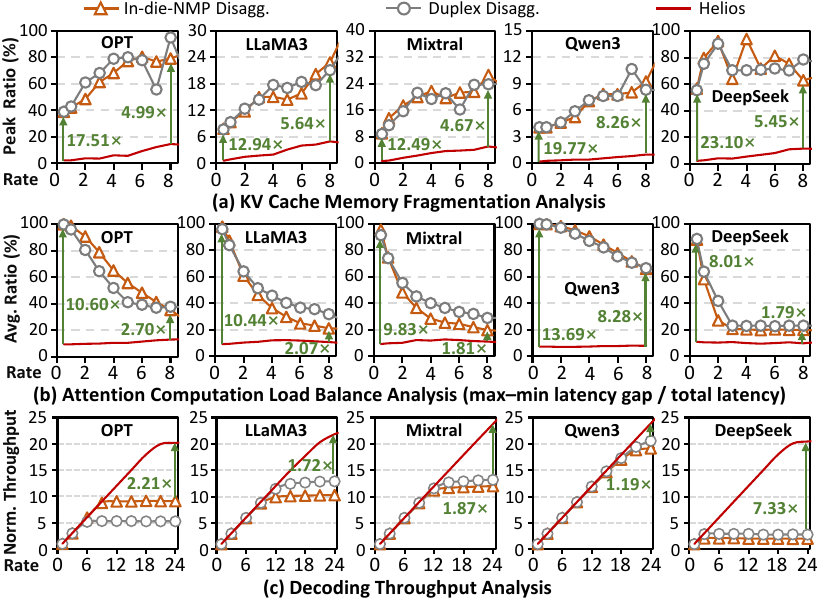} 
    \vspace{-1.75em}
    \caption{Attention Management Comparison between \TitleAbbr~and Prior NMPs.}
    \vspace{-1.25em}
    \label{fig:attn-lb}
\end{figure}

\noindent\textbf{Memory Fragmentation Analysis:}
To evaluate KV cache memory fragmentation between \TitleAbbr~and existing NMP designs, we measure the peak inter-PE/cube memory usage disparity under decoding serving stress tests, which is normalized to the available capacity of one PE/cube.
As shown in Fig.~\ref{fig:attn-lb}-(a), NMP baselines incur 4.67-23.10× peak fragmentation ratio than \TitleAbbr. 
As the load rises, NMP baselines exhibit a 1.66-3.77× escalation in peak fragmentation ratio.
On OPT/DeepSeek, this ratio even reaches \textasciitilde100\% of each cube's max capacity.
These results indicate that coarse-grained schemes cannot accommodate highly dynamic serving workloads.
In contrast, \TitleAbbr's fine-grained strategy maintains peak fragmentation within 0.2-15.0\%, enabling HB-Device to sustain high performance across all loads.

\noindent\textbf{Load Balance Analysis:}
We then assess attention computation's load balance by profiling the average ratio of max inter-PE/cube latency gap in total attention latency under serving stress tests.
As shown in Fig.~\ref{fig:attn-lb}-(b), NMP baselines incur 1.79-13.69× higher imbalance ratio than \TitleAbbr.
At low load, the batch size is too small to occupy all cubes in NMP baselines, causing imbalance ratio to reach nearly 100\%.
Although this ratio falls when request rate rises, it still remains at 19.9-66.6\%.
In contrast, fine-grained cache management and \textit{spatially-aware} block allocation enable \TitleAbbr~to confine this ratio within 7.1-13.3\%, thus fully utilizing the compute resources across highly dynamic serving workloads.

\noindent\textbf{Serving Capacity Analysis:}
To more intuitively illustrate the serving capacity benefits of \TitleAbbr's dynamic KV cache management, we compare decoding throughput across different designs.
As shown in Fig.~\ref{fig:attn-lb}-(c), \TitleAbbr~achieves 1.19×-7.33× higher throughput than prior NMP solutions.
Moreover, prior NMP designs saturate at lower request arrival rates.
These results confirm that \TitleAbbr~significantly improves the serving capacity compared with existing NMP architectures.

\begin{figure}
    \centering
    \includegraphics[width=0.99\columnwidth]{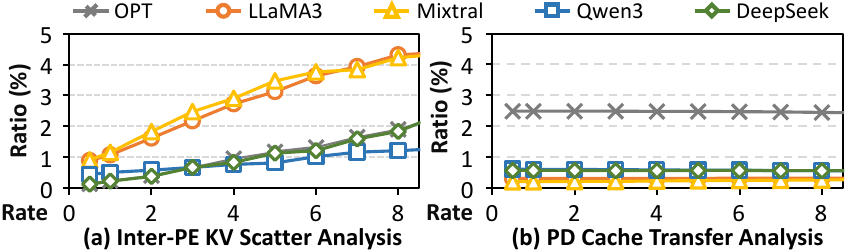}
    \vspace{-1.5em}
    \caption{KV Cache Transfer Overhead Analysis.}
     \vspace{-0.5em}
    \label{fig:kv-comm-analysis}
\end{figure}

\noindent\textbf{Cache Transfer Overhead:}
We begin by measuring the average fraction of scatter transfers
(Fig.~\ref{fig:qkv-fc-comm}-(b)) in total query/key/value projection latency under stress tests.
As illustrated in Fig.~\ref{fig:kv-comm-analysis}-(a), the proportion remains $\le$4.4\%, proving the capability of  Algorithm~\ref{algo:block-allocation} in optimizing inter-PE transfer overhead across varying load pressures.
We then profile the ratio of non-overlapped KV cache transfer overhead between prefill and decoding instances in E2E serving.
As shown in Fig.~\ref{fig:kv-comm-analysis}-(b), it remains $\le$2.5\%, 
demonstrating the efficiency of the cache transfer mechanism in Fig.~\ref{fig:cache-transfer-overlap}.

\begin{figure}
    \centering
    \includegraphics[width=0.99\columnwidth]{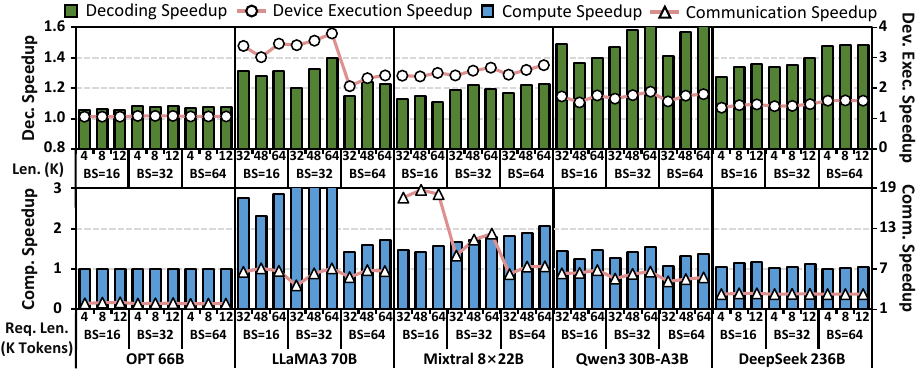}
    \vspace{-1.5em}
    \caption{Comparison with WaferLLM's Execution Flow.}
     \vspace{-0.5em}
    \label{fig:wafer-llm-comparison}
\end{figure}

\subsection{Comparison with WaferLLM}

A recent work, WaferLLM~\cite{waferllm0}, also designs decoding execution on mesh-based multi-PE accelerators.
Although it adopts similar FC execution to \TitleAbbr~ (Fig.~\ref{fig:attn-oproj-comm}-(b)), \textbf{\textit{similar to existing NMP designs in Table~\ref{tab:nmp-baseline}, it decomposes attention into independent GEMMs}}. This leads to two limitations:

\noindent\textbf{Low Compute Efficiency:}
\uline{\textit{First}}, 
The two attention GEMMs involve query-head, context-length, and head-dim dimensions.
Since query-head and head-dim are small (1-128 and 128-576), 
further partitioning them across the mesh, as in WaferLLM, reduces per-PE arithmetic intensity and results in small per-PE workloads, limiting compute utilization.
\uline{\textit{Second}}, as discussed in Sec.~\ref{sec:llm-serving}, context length varies at runtime due to request dynamicity. Therefore, it is difficult to adopt a unified intra-PE tiling strategy that maintains high efficiency across arbitrary context lengths.
In their latest implementation~\cite{waferllm0}, similar to existing NMP designs in Table~\ref{tab:nmp-baseline}, they assign a fixed per-PE context length for each request (seq\_len\_p\_pe in the code) to KV cache tensors~\cite{waferllm1,waferllm2,waferllm3,waferllm4}, which is pre-defined before execution~\cite{waferllm5,waferllm6}.
This requires us to preserve full context window for each request due to the request length dynamicity, leading to severe internal fragmentation as analyzed in Sec.~\ref{sec:nmp-limitation}.
\uline{\textit{In contrast}}, our tiled attention avoids partitioning along the short dimensions. By adjusting only the macro block size in Fig.~\ref{fig:on-chip-execution}-(b), we support a unified intra-PE tiling for arbitrary context lengths while preserving high efficiency.

\noindent\textbf{Low Inter-PE Communication Efficiency:}
\uline{\textit{First}}, WaferLLM executes softmax explicitly between the two attention GEMMs~\cite{waferllm9}, incurring additional overhead.
\uline{\textit{Second}}, its end-to-end decoding implementation~\cite{waferllm0} does not specify how newly generated KV vectors are inserted into the KV cache and only contains partial-sum all-reduce operators~\cite{waferllm7}.
Although a shift operation is proposed for this purpose, it is implemented isolatedly~\cite{waferllm8} and is not integrated into the end-to-end decoding pipeline~\cite{waferllm0}.
\uline{\textit{In contrast}}, our design overlaps softmax with attention GEMMs (Fig.~\ref{fig:on-chip-execution}-(b)) and provides an integrated execution flow for KV vector transfer (Fig.~\ref{fig:qkv-fc-comm}-(b)).

We implement WaferLLM’s execution flow~\cite{waferllm0} on HB-Device and compare it against \TitleAbbr~across multiple models. Workloads consist of variable-length requests randomly sampled from MoonCake dataset~\cite{qin2025mooncake} given the max context length.
To show WaferLLM's performance upper bound, we manually tune the optimal intra-PE tiling factors for different context lengths, and assume KV vector transfer incurs no overhead.
As shown in the first row of Fig.~\ref{fig:wafer-llm-comparison}, \TitleAbbr~achieves 1.06-1.68× decoding speedup and 1.07-3.81× device-execution speedup (excluding NVLink overhead) over WaferLLM.
We further analyze PE computation and inter-PE communication speedup.
As shown in the second row of Fig.~\ref{fig:wafer-llm-comparison}, although \TitleAbbr~only achieves 1.04×/2.51× average computation/communication speedup under short-context OPT/Deepseek workloads, the average speedup reaches 1.55×/6.11×  under other long-context workloads, confirming the superiority of our design against WaferLLM.

\subsection{Sensitivity Analysis}

\begin{figure}
    \centering
    \includegraphics[width=0.99\linewidth]{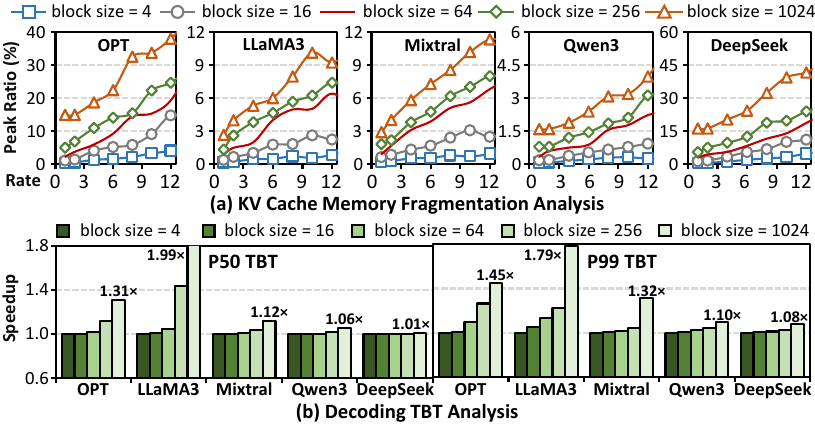} 
    \vspace{-2em}
    \caption{Performance Analysis of Different KV Block Sizes.}
    \vspace{-0.75em}
    \label{fig:kv-block-analysis}
\end{figure}

\noindent\textbf{KV Block Size:}
Fig.~\ref{fig:kv-block-analysis} analyzes the effect of KV block size.
For each model, we use the same block size across all layers and evaluate memory fragmentation and TBT with block sizes of 4/16/64/256/1024.
Since TBT distributions are similar across block sizes at low request rates, we report peak p50/p99 TBT speedups.
Increasing block size raises memory fragmentation, yet even with a block size of 1024, in-die NMP/Duplex exhibit 2.26-8.86× higher fragmentation than \TitleAbbr.
Besides, larger block sizes also improve TBT by enabling more contiguous DRAM accesses for KV cache.
Therefore, moderate block sizes (e.g., 64-256) can be used to balance such a fragmentation-TBT trade off, while larger block sizes (e.g., 1024) are preferable for latency-sensitive scenarios.
For inter-PE communication, block size only affects KV vector transfers. As Algorithm 1 minimizes this overhead (average 2.41\% in Fig.~\ref{fig:communication-breakdown}), its impact on communication is negligible.

\begin{figure}
    \centering
    \includegraphics[width=0.99\columnwidth]{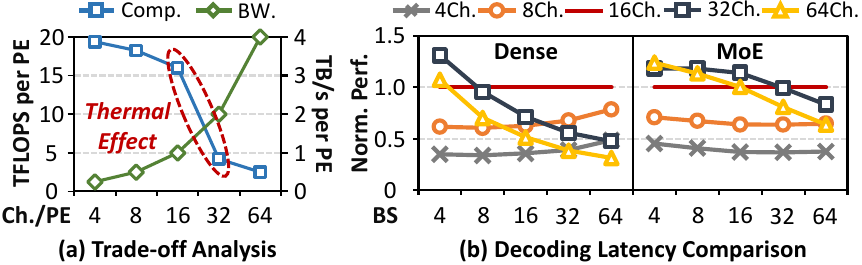}
    \vspace{-1.5em}
    \caption{Compute-Bandwidth Allocation Trade-off Analysis.}
     \vspace{-0.5em}
    \label{fig:comp-bw-analysis}
\end{figure}

\noindent\textbf{Compute-Bandwidth Allocation:}
3D-DRAM's memory controllers share the same logic-die area with compute logic.
Therefore, increasing bandwidth reduces the area for compute, creating a compute-bandwidth trade-off~\cite{li2025h2}.
To study this trade-off, we evaluate the allocable compute capacity under different DRAM channel counts, subject to the thermal constraints in Sec.~\ref{sec:ppa} and the 4×4 PE array of HB-Device.
As shown in Fig.~\ref{fig:comp-bw-analysis}-(a), reducing the channel count below the baseline 16 yields only marginal compute gains.
In contrast, increasing channel count significantly raises DRAM power.
Although area permits additional compute, thermal constraints lead to a substantial reduction in effective compute capacity.

Fig.~\ref{fig:comp-bw-analysis}-(b) compares decoding performance under different allocations.
For each batch size, we sweep context lengths from 4K to 128K (with 1K step) and average non-OOM workloads’ performance across dense and MoE models.
Reducing channel count yields only marginal compute gains, which are insufficient to offset the bandwidth-induced performance degradation.
Conversely, increasing channel count benefits small-batch cases—particularly for MoE models with expert routing—but significantly degrades performance at large batch sizes.
Therefore, allocating 4/8/32/64 channels achieves only 38.9\%/65.9\%/93.4\%/76.8\% of 16-channel's performance.

\section{Conclusion}

This paper proposes \TitleAbbr, a hybrid-bonding-based LLM serving accelerator with architecture-system co-design.
\TitleAbbr~efficiently supports \textit{NMP-native} distributed  tiled attention by customizing HB-Device architecture and execution flow, thus enabling \textit{NMP-native}  dynamic KV cache management.
\TitleAbbr~further customizes \textit{spatially-aware} KV cache allocation strategy, which not only balances the compute workload distribution, but also minimizes the inter-PE data transfer overhead under highly dynamic serving workloads.
Compared with existing GPU/NMP designs, \TitleAbbr~achieves 3.25× (geomean) speedup and 3.36× (geomean) better energy efficiency.


\bibliographystyle{IEEEtranS}
\bibliography{refs}

@inproceedings{fujun2020stacked,
  title={A stacked embedded DRAM array for LPDDR4/4X using hybrid bonding 3D integration with 34GB/s/1Gb 0.88 pJ/b logic-to-memory interface},
  author={Fujun, Bai and Xiping, Jiang and Song, Wang and Bing, Yu and Jie, Tan and Fengguo, Zuo and Chunjuan, Wang and Fan, Wang and Xiaodong, Long and Guoqing, Yu and Ni, Fu and Qiannan, Li and Hua, Li and Kexin, Wang and Huifu, Duan and Liang, Bai and Xuerong, Jia and Jin, Li and Mei, Li and Zhengwen, Wang and Sheng, Hu and Jun, Zhou and Qiong, Zhan and Peng, Sun and Daohong, Yang and Kau, Cheichan and Yang, David and Ho, Ching-Sung and Hongbin, Sun and Hangbing, Lv and Ming, Liu and Yi, Kang and Qiwei, Ren},
  booktitle={2020 IEEE International Electron Devices Meeting (IEDM)},
  pages={6--6},
  year={2020},
  organization={IEEE}
}

@inproceedings{niu2022184qps,
  title={184QPS/W 64Mb/mm 2 3D logic-to-DRAM hybrid bonding with process-near-memory engine for recommendation system},
  author={Dimin Niu and
                  Shuangchen Li and
                  Yuhao Wang and
                  Wei Han and
                  Zhe Zhang and
                  Yijin Guan and
                  Tianchan Guan and
                  Fei Sun and
                  Fei Xue and
                  Lide Duan and
                  Yuanwei Fang and
                  Hongzhong Zheng and
                  Xiping Jiang and
                  Song Wang and
                  Fengguo Zuo and
                  Yubing Wang and
                  Bing Yu and
                  Qiwei Ren and
                  Yuan Xie},
  booktitle={2022 IEEE International Solid-State Circuits Conference (ISSCC)},
  volume={65},
  pages={1--3},
  year={2022},
  organization={IEEE}
}

@inproceedings{wuu20223d,
  title={3D V-Cache: the Implementation of a Hybrid-Bonded 64MB Stacked Cache for a 7nm x86-64 CPU},
  author={Wuu, John and Agarwal, Rahul and Ciraula, Michael and Dietz, Carl and Johnson, Brett and Johnson, Dave and Schreiber, Russell and Swaminathan, Raja and Walker, Will and Naffziger, Samuel},
  booktitle={2022 IEEE International Solid-State Circuits Conference (ISSCC)},
  volume={65},
  pages={428--429},
  year={2022},
  organization={IEEE}
}

@inproceedings{yue2024exploiting,
  title={Exploiting similarity opportunities of emerging vision ai models on hybrid bonding architecture},
  author={Yue, Zhiheng and Wang, Huizheng and Fang, Jiahao and Deng, Jinyi and Lu, Guangyang and Tu, Fengbin and Guo, Ruiqi and Li, Yuxuan and Qin, Yubin and Wang, Yang and Li, Chao and Han, Huiming and Wei, Shaojun and Hu, Yang and Yin, Shouyi},
  booktitle={2024 ACM/IEEE 51st Annual International Symposium on Computer Architecture (ISCA)},
  pages={396--409},
  year={2024},
  organization={IEEE}
}

@inproceedings{wang2023135,
  title={A 135 GBps/Gbit 0.66 pJ/bit stacked embedded DRAM with multilayer arrays by fine pitch hybrid bonding and mini-TSV},
  author={Wang, Song and Yu, Bing and Xiao, Wenwu and Bai, Fujun and Long, Xiaodong and Bai, Liang and Jia, Xuerong and Zuo, Fengguo and Tan, Jie and Guo, Yixin and Sun, Peng and Zhou, Jun and Zhan, Qiong and Hu, Sheng and Zhou, Yu and Kang, Yi and Ren, Qiwei and Jiang, Xiping},
  booktitle={2023 IEEE Symposium on VLSI Technology and Circuits (VLSI Technology and Circuits)},
  pages={1--2},
  year={2023},
  organization={IEEE}
}

@article{wang20253d,
  title={A 3D Unified Analysis Method (3D-UAM) for Wafer-on-Wafer Stacked Near-Memory Structure},
  author={Wang, Song and Guo, Yixin and Tao, Wei and Jia, Xuerong and Bai, Fujun and Tan, Jie and Wang, Yubing and Bai, Liang and Guo, Fuzhi and Liu, Qi and Li, Jin and Yin, Peng and Liu, Fenning and Liu, Jing and Long, Xiaodong and Han, Yanwu and Yu, Zhongcheng and Cheng, Mengzi and Chen, Song and Jiang, Xiping},
  journal={IEEE Transactions on Very Large Scale Integration (VLSI) Systems},
  year={2025},
  publisher={IEEE}
}

@inproceedings{li2025h2,
  title={H2-LLM: Hardware-Dataflow Co-Exploration for Heterogeneous Hybrid-Bonding-based Low-Batch LLM Inference},
  author={Li, Cong and Yin, Yihan and Wu, Xintong and Zhu, Jingchen and Gao, Zhutianya and Niu, Dimin and Wu, Qiang and Si, Xin and Xie, Yuan and Zhang, Chen and Sun, Guangyu},
  booktitle={Proceedings of the 52nd Annual International Symposium on Computer Architecture},
  pages={194--210},
  year={2025}
}

@inproceedings{heo2024neupims,
  title={Neupims: Npu-pim heterogeneous acceleration for batched llm inferencing},
  author={Heo, Guseul and Lee, Sangyeop and Cho, Jaehong and Choi, Hyunmin and Lee, Sanghyeon and Ham, Hyungkyu and Kim, Gwangsun and Mahajan, Divya and Park, Jongse},
  booktitle={Proceedings of the 29th ACM International Conference on Architectural Support for Programming Languages and Operating Systems, Volume 3},
  pages={722--737},
  year={2024}
}

@inproceedings{kim2024sk,
  title={SK Hynix AI-Specific Computing Memory Solution: From AiM Device to Heterogeneous AiMX-xPU System for Comprehensive LLM Inference},
  author={Guhyun Kim and
                  Jinkwon Kim and
                  Nahsung Kim and
                  Woojae Shin and
                  Jongsoon Won and
                  Hyunha Joo and
                  Haerang Choi and
                  Byeongju An and
                  Gyeongcheol Shin and
                  Dayeon Yun and
                  Jeongbin Kim and
                  Changhyun Kim and
                  Ilkon Kim and
                  Jaehan Park and
                  Yosub Song and
                  Byeongsu Yang and
                  Hyeongdeok Lee and
                  Seungyeong Park and
                  Wonjun Lee and
                  Seonghun Kim and
                  Yonghoon Park and
                  Yousub Jung and
                  Gi{-}Ho Park and
                  Euicheol Lim},
  booktitle={2024 IEEE Hot Chips 36 Symposium (HCS)},
  pages={1--26},
  year={2024},
  organization={IEEE Computer Society}
}

@inproceedings{kim2023samsung,
  title={Samsung pim/pnm for transfmer based ai: Energy efficiency on pim/pnm cluster},
  author={Jin Hyun Kim and
                  Yuhwan Ro and
                  Jinin So and
                  Sukhan Lee and
                  Shinhaeng Kang and
                  YeonGon Cho and
                  Hyeonsu Kim and
                  Byeongho Kim and
                  Kyungsoo Kim and
                  Sangsoo Park and
                  Jin{-}Seong Kim and
                  Sanghoon Cha and
                  Won{-}Jo Lee and
                  Jin Jung and
                  Jonggeon Lee and
                  Jieun Lee and
                  Joon{-}Ho Song and
                  Seungwon Lee and
                  Jeonghyeon Cho and
                  Jaehoon Yu and
                  Kyomin Sohn},
  booktitle={2023 IEEE Hot Chips 35 Symposium (HCS)},
  pages={1--31},
  year={2023},
  organization={IEEE Computer Society}
}

@inproceedings{li2024specpim,
  title={Specpim: Accelerating speculative inference on pim-enabled system via architecture-dataflow co-exploration},
  author={Li, Cong and Zhou, Zhe and Zheng, Size and Zhang, Jiaxi and Liang, Yun and Sun, Guangyu},
  booktitle={Proceedings of the 29th ACM International Conference on Architectural Support for Programming Languages and Operating Systems, Volume 3},
  pages={950--965},
  year={2024}
}

@inproceedings{park2024attacc,
  title={Attacc! unleashing the power of pim for batched transformer-based generative model inference},
  author={Park, Jaehyun and Choi, Jaewan and Kyung, Kwanhee and Kim, Michael Jaemin and Kwon, Yongsuk and Kim, Nam Sung and Ahn, Jung Ho},
  booktitle={Proceedings of the 29th ACM International Conference on Architectural Support for Programming Languages and Operating Systems, Volume 2},
  pages={103--119},
  year={2024}
}

@inproceedings{yun2024duplex,
  title={Duplex: A device for large language models with mixture of experts, grouped query attention, and continuous batching},
  author={Yun, Sungmin and Kyung, Kwanhee and Cho, Juhwan and Choi, Jaewan and Kim, Jongmin and Kim, Byeongho and Lee, Sukhan and Sohn, Kyomin and Ahn, Jung Ho},
  booktitle={2024 57th IEEE/ACM International Symposium on Microarchitecture (MICRO)},
  pages={1429--1443},
  year={2024},
  organization={IEEE}
}

@inproceedings{he2025papi,
  title={Papi: Exploiting dynamic parallelism in large language model decoding with a processing-in-memory-enabled computing system},
  author={He, Yintao and Mao, Haiyu and Giannoula, Christina and Sadrosadati, Mohammad and G{\'o}mez-Luna, Juan and Li, Huawei and Li, Xiaowei and Wang, Ying and Mutlu, Onur},
  booktitle={Proceedings of the 30th ACM International Conference on Architectural Support for Programming Languages and Operating Systems, Volume 2},
  pages={766--782},
  year={2025}
}

@inproceedings{patel2024splitwise,
  title={Splitwise: Efficient generative llm inference using phase splitting},
  author={Patel, Pratyush and Choukse, Esha and Zhang, Chaojie and Shah, Aashaka and Goiri, {\'I}{\~n}igo and Maleki, Saeed and Bianchini, Ricardo},
  booktitle={2024 ACM/IEEE 51st Annual International Symposium on Computer Architecture (ISCA)},
  pages={118--132},
  year={2024},
  organization={IEEE}
}

@inproceedings{qin2025mooncake,
  title={Mooncake: Trading more storage for less computation—a $\{$KVCache-centric$\}$ architecture for serving $\{$LLM$\}$ chatbot},
  author={Qin, Ruoyu and Li, Zheming and He, Weiran and Cui, Jialei and Ren, Feng and Zhang, Mingxing and Wu, Yongwei and Zheng, Weimin and Xu, Xinran},
  booktitle={23rd USENIX Conference on File and Storage Technologies (FAST 25)},
  pages={155--170},
  year={2025}
}

@inproceedings{zhong2024distserve,
  title={$\{$DistServe$\}$: Disaggregating prefill and decoding for goodput-optimized large language model serving},
  author={Zhong, Yinmin and Liu, Shengyu and Chen, Junda and Hu, Jianbo and Zhu, Yibo and Liu, Xuanzhe and Jin, Xin and Zhang, Hao},
  booktitle={18th USENIX Symposium on Operating Systems Design and Implementation (OSDI 24)},
  pages={193--210},
  year={2024}
}

@misc{ualink,
  author = {Ultra Accelerator Link Consortium, Inc.},
  title = {UALink 200 Rev 1.0 Specification.},
  howpublished = {\url{https://ualinkconsortium.org/specification/}},
  year={2025},
}

@misc{cxl-doc,
  author = {CXL Consortium},
  title = {Compute Express Link Specification Revision 3.0.},
  howpublished = {\url{https://www.computeexpresslink.org/download-the-specification}},
  year={2025},
}

@misc{nvlink,
    author = {NVIDIA},
    title = {Build Semi-Custom AI Infrastructure | NVIDIA NVLink Fusion.},
    howpublished = {\url{https://www.nvidia.com/en-us/data-center/nvlink-fusion/}},
    year = {2025}
}

@misc{mooncacke-transfer-engine,
    author = {Kimi},
    title = {Mooncake Transfer Engine.},
    howpublished = {\url{https://github.com/kvcache-ai/Mooncake}},
    year = {2025}
}

@article{zhang2022opt,
  title={Opt: Open pre-trained transformer language models},
  author={Susan Zhang and Stephen Roller and Naman Goyal and Mikel Artetxe and Moya Chen and Shuohui Chen and Christopher Dewan and Mona Diab and Xian Li and Xi Victoria Lin and Todor Mihaylov and Myle Ott and Sam Shleifer and Kurt Shuster and Daniel Simig and Punit Singh Koura and Anjali Sridhar and Tianlu Wang and Luke Zettlemoyer},
  journal={arXiv preprint arXiv:2205.01068},
  year={2022}
}

@article{dubey2024llama,
  title={The llama 3 herd of models},
  author={Abhimanyu Dubey and Abhinav Jauhri and Abhinav Pandey and Abhishek Kadian and Ahmad Al-Dahle and Aiesha Letman and Akhil Mathur and Alan Schelten and Amy Yang and Angela Fan and Anirudh Goyal and Anthony Hartshorn and Aobo Yang and Archi Mitra and Archie Sravankumar and Artem Korenev and Arthur Hinsvark and Arun Rao and Aston Zhang and Aurelien Rodriguez and Austen Gregerson and Ava Spataru and Baptiste Roziere and Bethany Biron and Binh Tang and Bobbie Chern and Charlotte Caucheteux and Chaya Nayak and Chloe Bi and Chris Marra and Chris McConnell and Christian Keller and Christophe Touret and Chunyang Wu and Corinne Wong and Cristian Canton Ferrer and Cyrus Nikolaidis and Damien Allonsius and Daniel Song and Danielle Pintz and Danny Livshits and David Esiobu and Dhruv Choudhary and Dhruv Mahajan and Diego Garcia-Olano and Diego Perino and Dieuwke Hupkes and Egor Lakomkin and Ehab AlBadawy and Elina Lobanova and Emily Dinan and Eric Michael Smith and Filip Radenovic and Frank Zhang and Gabriel Synnaeve and Gabrielle Lee and Georgia Lewis Anderson and Graeme Nail and Gregoire Mialon and Guan Pang and Guillem Cucurell and Hailey Nguyen and Hannah Korevaar and Hu Xu and Hugo Touvron and Iliyan Zarov and Imanol Arrieta Ibarra and Isabel Kloumann and Ishan Misra and Ivan Evtimov and Jade Copet and Jaewon Lee and Jan Geffert and Jana Vranes and Jason Park and Jay Mahadeokar and Jeet Shah and Jelmer van der Linde and Jennifer Billock and Jenny Hong and Jenya Lee and Jeremy Fu and Jianfeng Chi and Jianyu Huang and Jiawen Liu and Jie Wang and Jiecao Yu and Joanna Bitton and Joe Spisak and Jongsoo Park and Joseph Rocca and Joshua Johnstun and Joshua Saxe and Junteng Jia and Kalyan Vasuden Alwala and Kartikeya Upasani and Kate Plawiak and Ke Li and Kenneth Heafield and Kevin Stone and Khalid El-Arini and Krithika Iyer and Kshitiz Malik and Kuenley Chiu and Kunal Bhalla and Lauren Rantala-Yeary and Laurens van der Maaten and Lawrence Chen and Liang Tan and Liz Jenkins and Louis Martin and Lovish Madaan and Lubo Malo and Lukas Blecher and Lukas Landzaat and Luke de Oliveira and Madeline Muzzi and Mahesh Pasupuleti and Mannat Singh and Manohar Paluri and Marcin Kardas and Mathew Oldham and Mathieu Rita and Maya Pavlova and Melanie Kambadur and Mike Lewis and Min Si and Mitesh Kumar Singh and Mona Hassan and Naman Goyal and Narjes Torabi and Nikolay Bashlykov and Nikolay Bogoychev and Niladri Chatterji and Olivier Duchenne and Onur Çelebi and Patrick Alrassy and Pengchuan Zhang and Pengwei Li and Petar Vasic and Peter Weng and Prajjwal Bhargava and Pratik Dubal and Praveen Krishnan and Punit Singh Koura and Puxin Xu and Qing He and Qingxiao Dong and Ragavan Srinivasan and Raj Ganapathy and Ramon Calderer and Ricardo Silveira Cabral and Robert Stojnic and Roberta Raileanu and Rohit Girdhar and Rohit Patel and Romain Sauvestre and Ronnie Polidoro and Roshan Sumbaly and Ross Taylor and Ruan Silva and Rui Hou and Rui Wang and Saghar Hosseini and Sahana Chennabasappa and Sanjay Singh and Sean Bell and Seohyun Sonia Kim and Sergey Edunov and Shaoliang Nie and Sharan Narang and Sharath Raparthy and Sheng Shen and Shengye Wan and Shruti Bhosale and Shun Zhang and Simon Vandenhende and Soumya Batra and Spencer Whitman and Sten Sootla and Stephane Collot and Suchin Gururangan and Sydney Borodinsky and Tamar Herman and Tara Fowler and Tarek Sheasha and Thomas Georgiou and Thomas Scialom and Tobias Speckbacher and Todor Mihaylov and Tong Xiao and Ujjwal Karn and Vedanuj Goswami and Vibhor Gupta and Vignesh Ramanathan and Viktor Kerkez and Vincent Gonguet and Virginie Do and Vish Vogeti and Vladan Petrovic and Weiwei Chu and Wenhan Xiong and Wenyin Fu and Whitney Meers and Xavier Martinet and Xiaodong Wang and Xiaoqing Ellen Tan and Xinfeng Xie and Xuchao Jia and Xuewei Wang and Yaelle Goldschlag and Yashesh Gaur and Yasmine Babaei and Yi Wen and Yiwen Song and Yuchen Zhang and Yue Li and Yuning Mao and Zacharie Delpierre Coudert and Zheng Yan and Zhengxing Chen and Zoe Papakipos and Aaditya Singh and Aaron Grattafiori and Abha Jain and Adam Kelsey and Adam Shajnfeld and Adithya Gangidi and Adolfo Victoria and Ahuva Goldstand and Ajay Menon and Ajay Sharma and Alex Boesenberg and Alex Vaughan and Alexei Baevski and Allie Feinstein and Amanda Kallet and Amit Sangani and Anam Yunus and Andrei Lupu and Andres Alvarado and Andrew Caples and Andrew Gu and Andrew Ho and Andrew Poulton and Andrew Ryan and Ankit Ramchandani and Annie Franco and Aparajita Saraf and Arkabandhu Chowdhury and Ashley Gabriel and Ashwin Bharambe and Assaf Eisenman and Azadeh Yazdan and Beau James and Ben Maurer and Benjamin Leonhardi and Bernie Huang and Beth Loyd and Beto De Paola and Bhargavi Paranjape and Bing Liu and Bo Wu and Boyu Ni and Braden Hancock and Bram Wasti and Brandon Spence and Brani Stojkovic and Brian Gamido and Britt Montalvo and Carl Parker and Carly Burton and Catalina Mejia and Changhan Wang and Changkyu Kim and Chao Zhou and Chester Hu and Ching-Hsiang Chu and Chris Cai and Chris Tindal and Christoph Feichtenhofer and Damon Civin and Dana Beaty and Daniel Kreymer and Daniel Li and Danny Wyatt and David Adkins and David Xu and Davide Testuggine and Delia David and Devi Parikh and Diana Liskovich and Didem Foss and Dingkang Wang and Duc Le and Dustin Holland and Edward Dowling and Eissa Jamil and Elaine Montgomery and Eleonora Presani and Emily Hahn and Emily Wood and Erik Brinkman and Esteban Arcaute and Evan Dunbar and Evan Smothers and Fei Sun and Felix Kreuk and Feng Tian and Firat Ozgenel and Francesco Caggioni and Francisco Guzmán and Frank Kanayet and Frank Seide and Gabriela Medina Florez and Gabriella Schwarz and Gada Badeer and Georgia Swee and Gil Halpern and Govind Thattai and Grant Herman and Grigory Sizov and Guangyi and Zhang and Guna Lakshminarayanan and Hamid Shojanazeri and Han Zou and Hannah Wang and Hanwen Zha and Haroun Habeeb and Harrison Rudolph and Helen Suk and Henry Aspegren and Hunter Goldman and Ibrahim Damlaj and Igor Molybog and Igor Tufanov and Irina-Elena Veliche and Itai Gat and Jake Weissman and James Geboski and James Kohli and Japhet Asher and Jean-Baptiste Gaya and Jeff Marcus and Jeff Tang and Jennifer Chan and Jenny Zhen and Jeremy Reizenstein and Jeremy Teboul and Jessica Zhong and Jian Jin and Jingyi Yang and Joe Cummings and Jon Carvill and Jon Shepard and Jonathan McPhie and Jonathan Torres and Josh Ginsburg and Junjie Wang and Kai Wu and Kam Hou U and Karan Saxena and Karthik Prasad and Kartikay Khandelwal and Katayoun Zand and Kathy Matosich and Kaushik Veeraraghavan and Kelly Michelena and Keqian Li and Kun Huang and Kunal Chawla and Kushal Lakhotia and Kyle Huang and Lailin Chen and Lakshya Garg and Lavender A and Leandro Silva and Lee Bell and Lei Zhang and Liangpeng Guo and Licheng Yu and Liron Moshkovich and Luca Wehrstedt and Madian Khabsa and Manav Avalani and Manish Bhatt and Maria Tsimpoukelli and Martynas Mankus and Matan Hasson and Matthew Lennie and Matthias Reso and Maxim Groshev and Maxim Naumov and Maya Lathi and Meghan Keneally and Michael L. Seltzer and Michal Valko and Michelle Restrepo and Mihir Patel and Mik Vyatskov and Mikayel Samvelyan and Mike Clark and Mike Macey and Mike Wang and Miquel Jubert Hermoso and Mo Metanat and Mohammad Rastegari and Munish Bansal and Nandhini Santhanam and Natascha Parks and Natasha White and Navyata Bawa and Nayan Singhal and Nick Egebo and Nicolas Usunier and Nikolay Pavlovich Laptev and Ning Dong and Ning Zhang and Norman Cheng and Oleg Chernoguz and Olivia Hart and Omkar Salpekar and Ozlem Kalinli and Parkin Kent and Parth Parekh and Paul Saab and Pavan Balaji and Pedro Rittner and Philip Bontrager and Pierre Roux and Piotr Dollar and Polina Zvyagina and Prashant Ratanchandani and Pritish Yuvraj and Qian Liang and Rachad Alao and Rachel Rodriguez and Rafi Ayub and Raghotham Murthy and Raghu Nayani and Rahul Mitra and Raymond Li and Rebekkah Hogan and Robin Battey and Rocky Wang and Rohan Maheswari and Russ Howes and Ruty Rinott and Sai Jayesh Bondu and Samyak Datta and Sara Chugh and Sara Hunt and Sargun Dhillon and Sasha Sidorov and Satadru Pan and Saurabh Verma and Seiji Yamamoto and Sharadh Ramaswamy and Shaun Lindsay and Shaun Lindsay and Sheng Feng and Shenghao Lin and Shengxin Cindy Zha and Shiva Shankar and Shuqiang Zhang and Shuqiang Zhang and Sinong Wang and Sneha Agarwal and Soji Sajuyigbe and Soumith Chintala and Stephanie Max and Stephen Chen and Steve Kehoe and Steve Satterfield and Sudarshan Govindaprasad and Sumit Gupta and Sungmin Cho and Sunny Virk and Suraj Subramanian and Sy Choudhury and Sydney Goldman and Tal Remez and Tamar Glaser and Tamara Best and Thilo Kohler and Thomas Robinson and Tianhe Li and Tianjun Zhang and Tim Matthews and Timothy Chou and Tzook Shaked and Varun Vontimitta and Victoria Ajayi and Victoria Montanez and Vijai Mohan and Vinay Satish Kumar and Vishal Mangla and Vítor Albiero and Vlad Ionescu and Vlad Poenaru and Vlad Tiberiu Mihailescu and Vladimir Ivanov and Wei Li and Wenchen Wang and Wenwen Jiang and Wes Bouaziz and Will Constable and Xiaocheng Tang and Xiaofang Wang and Xiaojian Wu and Xiaolan Wang and Xide Xia and Xilun Wu and Xinbo Gao and Yanjun Chen and Ye Hu and Ye Jia and Ye Qi and Yenda Li and Yilin Zhang and Ying Zhang and Yossi Adi and Youngjin Nam and Yu and Wang and Yuchen Hao and Yundi Qian and Yuzi He and Zach Rait and Zachary DeVito and Zef Rosnbrick and Zhaoduo Wen and Zhenyu Yang and Zhiwei Zhao},
  journal={arXiv preprint arXiv:2407.21783},
  year={2024}
}

@article{jiang2024mixtral,
  title={Mixtral of experts},
  author={Albert Q. Jiang and Alexandre Sablayrolles and Antoine Roux and Arthur Mensch and Blanche Savary and Chris Bamford and Devendra Singh Chaplot and Diego de las Casas and Emma Bou Hanna and Florian Bressand and Gianna Lengyel and Guillaume Bour and Guillaume Lample and Lélio Renard Lavaud and Lucile Saulnier and Marie-Anne Lachaux and Pierre Stock and Sandeep Subramanian and Sophia Yang and Szymon Antoniak and Teven Le Scao and Théophile Gervet and Thibaut Lavril and Thomas Wang and Timothée Lacroix and William El Sayed},
  journal={arXiv preprint arXiv:2401.04088},
  year={2024}
}

@article{liu2024deepseek,
  title={Deepseek-v2: A strong, economical, and efficient mixture-of-experts language model},
  author={DeepSeek-AI and Aixin Liu and Bei Feng and Bin Wang and Bingxuan Wang and Bo Liu and Chenggang Zhao and Chengqi Dengr and Chong Ruan and Damai Dai and Daya Guo and Dejian Yang and Deli Chen and Dongjie Ji and Erhang Li and Fangyun Lin and Fuli Luo and Guangbo Hao and Guanting Chen and Guowei Li and H. Zhang and Hanwei Xu and Hao Yang and Haowei Zhang and Honghui Ding and Huajian Xin and Huazuo Gao and Hui Li and Hui Qu and J. L. Cai and Jian Liang and Jianzhong Guo and Jiaqi Ni and Jiashi Li and Jin Chen and Jingyang Yuan and Junjie Qiu and Junxiao Song and Kai Dong and Kaige Gao and Kang Guan and Lean Wang and Lecong Zhang and Lei Xu and Leyi Xia and Liang Zhao and Liyue Zhang and Meng Li and Miaojun Wang and Mingchuan Zhang and Minghua Zhang and Minghui Tang and Mingming Li and Ning Tian and Panpan Huang and Peiyi Wang and Peng Zhang and Qihao Zhu and Qinyu Chen and Qiushi Du and R. J. Chen and R. L. Jin and Ruiqi Ge and Ruizhe Pan and Runxin Xu and Ruyi Chen and S. S. Li and Shanghao Lu and Shangyan Zhou and Shanhuang Chen and Shaoqing Wu and Shengfeng Ye and Shirong Ma and Shiyu Wang and Shuang Zhou and Shuiping Yu and Shunfeng Zhou and Size Zheng and T. Wang and Tian Pei and Tian Yuan and Tianyu Sun and W. L. Xiao and Wangding Zeng and Wei An and Wen Liu and Wenfeng Liang and Wenjun Gao and Wentao Zhang and X. Q. Li and Xiangyue Jin and Xianzu Wang and Xiao Bi and Xiaodong Liu and Xiaohan Wang and Xiaojin Shen and Xiaokang Chen and Xiaosha Chen and Xiaotao Nie and Xiaowen Sun and Xiaoxiang Wang and Xin Liu and Xin Xie and Xingkai Yu and Xinnan Song and Xinyi Zhou and Xinyu Yang and Xuan Lu and Xuecheng Su and Y. Wu and Y. K. Li and Y. X. Wei and Y. X. Zhu and Yanhong Xu and Yanping Huang and Yao Li and Yao Zhao and Yaofeng Sun and Yaohui Li and Yaohui Wang and Yi Zheng and Yichao Zhang and Yiliang Xiong and Yilong Zhao and Ying He and Ying Tang and Yishi Piao and Yixin Dong and Yixuan Tan and Yiyuan Liu and Yongji Wang and Yongqiang Guo and Yuchen Zhu and Yuduan Wang and Yuheng Zou and Yukun Zha and Yunxian Ma and Yuting Yan and Yuxiang You and Yuxuan Liu and Z. Z. Ren and Zehui Ren and Zhangli Sha and Zhe Fu and Zhen Huang and Zhen Zhang and Zhenda Xie and Zhewen Hao and Zhihong Shao and Zhiniu Wen and Zhipeng Xu and Zhongyu Zhang and Zhuoshu Li and Zihan Wang and Zihui Gu and Zilin Li and Ziwei Xie},
  journal={arXiv preprint arXiv:2405.04434},
  year={2024}
}

@article{yang2025qwen3,
  title={Qwen3 technical report},
  author={An Yang and Anfeng Li and Baosong Yang and Beichen Zhang and Binyuan Hui and Bo Zheng and Bowen Yu and Chang Gao and Chengen Huang and Chenxu Lv and Chujie Zheng and Dayiheng Liu and Fan Zhou and Fei Huang and Feng Hu and Hao Ge and Haoran Wei and Huan Lin and Jialong Tang and Jian Yang and Jianhong Tu and Jianwei Zhang and Jianxin Yang and Jiaxi Yang and Jing Zhou and Jingren Zhou and Junyang Lin and Kai Dang and Keqin Bao and Kexin Yang and Le Yu and Lianghao Deng and Mei Li and Mingfeng Xue and Mingze Li and Pei Zhang and Peng Wang and Qin Zhu and Rui Men and Ruize Gao and Shixuan Liu and Shuang Luo and Tianhao Li and Tianyi Tang and Wenbiao Yin and Xingzhang Ren and Xinyu Wang and Xinyu Zhang and Xuancheng Ren and Yang Fan and Yang Su and Yichang Zhang and Yinger Zhang and Yu Wan and Yuqiong Liu and Zekun Wang and Zeyu Cui and Zhenru Zhang and Zhipeng Zhou and Zihan Qiu},
  journal={arXiv preprint arXiv:2505.09388},
  year={2025}
}

@inproceedings{kwon2023efficient,
  title={Efficient memory management for large language model serving with pagedattention},
  author={Kwon, Woosuk and Li, Zhuohan and Zhuang, Siyuan and Sheng, Ying and Zheng, Lianmin and Yu, Cody Hao and Gonzalez, Joseph and Zhang, Hao and Stoica, Ion},
  booktitle={Proceedings of the 29th Symposium on Operating Systems Principles},
  pages={611--626},
  year={2023}
}

@article{kwon2020maestro,
  title={Maestro: A data-centric approach to understand reuse, performance, and hardware cost of dnn mappings},
  author={Kwon, Hyoukjun and Chatarasi, Prasanth and Sarkar, Vivek and Krishna, Tushar and Pellauer, Michael and Parashar, Angshuman},
  journal={IEEE micro},
  volume={40},
  number={3},
  pages={20--29},
  year={2020},
  publisher={IEEE}
}

@inproceedings{parashar2019timeloop,
  title={Timeloop: A systematic approach to dnn accelerator evaluation},
  author={Parashar, Angshuman and Raina, Priyanka and Shao, Yakun Sophia and Chen, Yu-Hsin and Ying, Victor A and Mukkara, Anurag and Venkatesan, Rangharajan and Khailany, Brucek and Keckler, Stephen W and Emer, Joel},
  booktitle={2019 IEEE international symposium on performance analysis of systems and software (ISPASS)},
  pages={304--315},
  year={2019},
  organization={IEEE}
}

@inproceedings{lu2021tenet,
  title={Tenet: A framework for modeling tensor dataflow based on relation-centric notation},
  author={Lu, Liqiang and Guan, Naiqing and Wang, Yuyue and Jia, Liancheng and Luo, Zizhang and Yin, Jieming and Cong, Jason and Liang, Yun},
  booktitle={2021 ACM/IEEE 48th Annual International Symposium on Computer Architecture (ISCA)},
  pages={720--733},
  year={2021},
  organization={IEEE}
}

@inproceedings{yu2022orca,
  title={Orca: A distributed serving system for $\{$Transformer-Based$\}$ generative models},
  author={Yu, Gyeong-In and Jeong, Joo Seong and Kim, Geon-Woo and Kim, Soojeong and Chun, Byung-Gon},
  booktitle={16th USENIX Symposium on Operating Systems Design and Implementation (OSDI 22)},
  pages={521--538},
  year={2022}
}

@misc{tile-norm,
author = {Chan, Tony F. and Golub, Gene H. and LeVeque, Randall J.},
title = {Updating formulae and a pairwise algorithm for computing sample variances},
year = {1979},
publisher = {Stanford University},
address = {Stanford, CA, USA},
}

@misc{ba2016layernormalization,
      title={Layer Normalization}, 
      author={Jimmy Lei Ba and Jamie Ryan Kiros and Geoffrey E. Hinton},
      year={2016},
      eprint={1607.06450},
      archivePrefix={arXiv},
      primaryClass={stat.ML},
      url={https://arxiv.org/abs/1607.06450}, 
}

@misc{zhang2019rootmeansquarelayer,
      title={Root Mean Square Layer Normalization}, 
      author={Biao Zhang and Rico Sennrich},
      year={2019},
      eprint={1910.07467},
      archivePrefix={arXiv},
      primaryClass={cs.LG},
      url={https://arxiv.org/abs/1910.07467}, 
}

@inproceedings{tidalmesh,
  author       = {Dongkyun Lim and
                  John Kim},
  title        = {TidalMesh: Topology-Driven AllReduce Collective Communication for
                  Mesh Topology},
  booktitle    = {{IEEE} International Symposium on High Performance Computer Architecture,
                  {HPCA} 2025, Las Vegas, NV, USA, March 1-5, 2025},
  pages        = {1526--1540},
  publisher    = {{IEEE}},
  year         = {2025},
}

@article{cctheory,
  author       = {Ernie Chan and
                  Marcel Heimlich and
                  Avi Purkayastha and
                  Robert A. van de Geijn},
  title        = {Collective communication: theory, practice, and experience},
  journal      = {Concurr. Comput. Pract. Exp.},
  volume       = {19},
  number       = {13},
  pages        = {1749--1783},
  year         = {2007},
}

@misc{milakov2018onlinenormalizercalculationsoftmax,
      title={Online normalizer calculation for softmax}, 
      author={Maxim Milakov and Natalia Gimelshein},
      year={2018},
      eprint={1805.02867},
      archivePrefix={arXiv},
      primaryClass={cs.PF},
      url={https://arxiv.org/abs/1805.02867}, 
}

@article{dao2022flashattention,
  title={Flashattention: Fast and memory-efficient exact attention with io-awareness},
  author={Dao, Tri and Fu, Dan and Ermon, Stefano and Rudra, Atri and R{\'e}, Christopher},
  journal={Advances in neural information processing systems},
  volume={35},
  pages={16344--16359},
  year={2022}
}

@article{dao2023flashattention,
  title={Flashattention-2: Faster attention with better parallelism and work partitioning},
  author={Dao, Tri},
  journal={arXiv preprint arXiv:2307.08691},
  year={2023}
}

@misc{flashdecoding,
    author = {Tri Dao and Daniel Haziza and Francisco Massa and Grigory Sizov},
    title = { Flash-decoding for long-context inference.},
    howpublished = {\url{https://crfm.stanford.edu/2023/10/12/flashdecoding.html}},
    year={2023}
}

@misc{shah2024flashattention3fastaccurateattention,
      title={FlashAttention-3: Fast and Accurate Attention with Asynchrony and Low-precision}, 
      author={Jay Shah and Ganesh Bikshandi and Ying Zhang and Vijay Thakkar and Pradeep Ramani and Tri Dao},
      year={2024},
      eprint={2407.08608},
      archivePrefix={arXiv},
      primaryClass={cs.LG},
      url={https://arxiv.org/abs/2407.08608}, 
}

@article{ye2025flashinfer,
    title = {FlashInfer: Efficient and Customizable Attention Engine for LLM Inference Serving},
    author = {
      Ye, Zihao and
      Chen, Lequn and
      Lai, Ruihang and
      Lin, Wuwei and
      Zhang, Yineng and
      Wang, Stephanie and
      Chen, Tianqi and
      Kasikci, Baris and
      Grover, Vinod and
      Krishnamurthy, Arvind and
      Ceze, Luis
    },
    journal = {arXiv preprint arXiv:2501.01005},
    year = {2025},
    url = {https://arxiv.org/abs/2501.01005}
}

@inproceedings{sglang,
  author       = {Lianmin Zheng and
                  Liangsheng Yin and
                  Zhiqiang Xie and
                  Chuyue Sun and
                  Jeff Huang and
                  Cody Hao Yu and
                  Shiyi Cao and
                  Christos Kozyrakis and
                  Ion Stoica and
                  Joseph E. Gonzalez and
                  Clark W. Barrett and
                  Ying Sheng},
  title        = {SGLang: Efficient Execution of Structured Language Model Programs},
  booktitle    = {Advances in Neural Information Processing Systems 38: Annual Conference
                  on Neural Information Processing Systems 2024, NeurIPS 2024, Vancouver,
                  BC, Canada, December 10 - 15, 2024},
  year         = {2024},
}

@article{luo2023ramulator,
  title={Ramulator 2.0: A Modern, Modular, and Extensible DRAM Simulator},
  author={Luo, Haocong and Tu, Yahya Can and Bostanc{\i}, F Nisa and Olgun, Ataberk and Ya, A Giray and Mutlu, Onur},
  journal={IEEE Computer Architecture Letters},
  year={2023},
  publisher={IEEE}
}

@misc{pyNVML,
    author = {Rjzamora},
    title = {pyNVML},
    howpublished = {\url{https://pypi.org/project/pynvml}},
    year={2024},
}

@misc{attacc-sim,
    author = {Jaehyun Park, Jaewan Choi},
    title = {Simulator for AttAcc},
    howpublished = {\url{https://github.com/scale-snu/attacc_simulator/}},
    year={2024},
}

@misc{duplex-sim,
    author = {Sungmin Yun, Kwanhee Kyung, Juhwan Cho},
    title = {LLMSimulator},
    howpublished = {\url{https://github.com/scale-snu/LLMSimulator/}},
    year={2025},
}

@inproceedings{nvlink-c2c,
  author       = {Ying Wei and
                  Yi Chieh Huang and
                  Haiming Tang and
                  Nithya Sankaran and
                  Ish Chadha and
                  Dai Dai and
                  Olakanmi Oluwole and
                  Vishnu Balan and
                  Edward Lee},
  title        = {NVLink-C2C: {A} Coherent Off Package Chip-to-Chip Interconnect with
                  40Gbps/pin Single-ended Signaling},
  booktitle    = {{IEEE} International Solid- State Circuits Conference, {ISSCC} 2023,
                  San Francisco, CA, USA, February 19-23, 2023},
  pages        = {160--161},
  publisher    = {{IEEE}},
  year         = {2023},
}

@misc{simdistserve,
    author = {Yinmin Zhong, Shengyu Liu},
    title = {DistServe Simulator},
    howpublished = {\url{https://github.com/LLMServe/DistServe/tree/main/simdistserve}},
    year={2024},
}

@inproceedings{jiang2013detailed,
  title={A detailed and flexible cycle-accurate network-on-chip simulator},
  author={Jiang, Nan and Becker, Daniel U and Michelogiannakis, George and Balfour, James and Towles, Brian and Shaw, David E and Kim, John and Dally, William J},
  booktitle={2013 IEEE international symposium on performance analysis of systems and software (ISPASS)},
  pages={86--96},
  year={2013},
  organization={IEEE}
}

@article{vaswani2017attention,
  title={Attention is all you need},
  author       = {Ashish Vaswani and
                  Noam Shazeer and
                  Niki Parmar and
                  Jakob Uszkoreit and
                  Llion Jones and
                  Aidan N. Gomez and
                  Lukasz Kaiser and
                  Illia Polosukhin},
  journal={Advances in Neural Information Processing Systems},
  year={2017}
}

@article{shazeer1911fast,
  title={Fast transformer decoding: One write-head is all you need, 2019},
  author={Shazeer, Noam},
  journal={URL https://arxiv. org/abs},
  pages={23},
  year={2019}
}

@article{ainslie2023gqa,
  title={Gqa: Training generalized multi-query transformer models from multi-head checkpoints},
  author={Ainslie, Joshua and Lee-Thorp, James and de Jong, Michiel and Zemlyanskiy, Yury and Lebr{\'o}n, Federico and Sanghai, Sumit},
  journal={arXiv preprint arXiv:2305.13245},
  year={2023}
}

@misc{deepseekai2025deepseekr1incentivizingreasoningcapability,
      title={DeepSeek-R1: Incentivizing Reasoning Capability in LLMs via Reinforcement Learning}, 
      author={DeepSeek-AI},
      year={2025},
      eprint={2501.12948},
      archivePrefix={arXiv},
      primaryClass={cs.CL},
      url={https://arxiv.org/abs/2501.12948}, 
}

@misc{deepseekai2024deepseekv3technicalreport,
      title={DeepSeek-V3 Technical Report}, 
      author={DeepSeek-AI},
      year={2024},
      eprint={2412.19437},
      archivePrefix={arXiv},
      primaryClass={cs.CL},
      url={https://arxiv.org/abs/2412.19437}, 
}

@article{shazeer2020glu,
  title={Glu variants improve transformer},
  author={Shazeer, Noam},
  journal={arXiv preprint arXiv:2002.05202},
  year={2020}
}

@misc{chatgpt,
    author = {OpenAI},
    title = {Chatgpt.},
    howpublished = {\url{https://chatgpt.com/}},
    year = {2025}
}

@misc{gemini,
    author = {Google},
    title = {Gemini.},
    howpublished = {\url{https://gemini.google.com/app}},
    year = {2025}
}

@misc{claude,
    author = {Anthropic},
    title = {Claude.},
    howpublished = {\url{https://www.anthropic.com/claude}},
    year = {2025}
}

@misc{deepseek,
    author = {DeepSeek AI},
    title = {DeepSeek.},
    howpublished = {\url{https://chat.deepseek.com/}},
    year = {2025}
}

@article{wang2024towards,
  title={Towards efficient and reliable llm serving: A real-world workload study},
  author={Wang, Yuxin and Chen, Yuhan and Li, Zeyu and Tang, Zhenheng and Guo, Rui and Wang, Xin and Wang, Qiang and Zhou, Amelie Chi and Chu, Xiaowen},
  journal={CoRR},
  year={2024}
}

@article{xiang2025servegen,
  title={ServeGen: Workload Characterization and Generation of Large Language Model Serving in Production},
  author={Xiang, Yuxing and Li, Xue and Qian, Kun and Yu, Wenyuan and Zhai, Ennan and Jin, Xin},
  journal={arXiv preprint arXiv:2505.09999},
  year={2025}
}

@misc{zhao2025insightsdeepseekv3scalingchallenges,
      title={Insights into DeepSeek-V3: Scaling Challenges and Reflections on Hardware for AI Architectures}, 
      author={Chenggang Zhao and Chengqi Deng and Chong Ruan and Damai Dai and Huazuo Gao and Jiashi Li and Liyue Zhang and Panpan Huang and Shangyan Zhou and Shirong Ma and Wenfeng Liang and Ying He and Yuqing Wang and Yuxuan Liu and Y. X. Wei},
      year={2025},
      eprint={2505.09343},
      archivePrefix={arXiv},
      primaryClass={cs.DC},
      url={https://arxiv.org/abs/2505.09343}, 
}

@article{nijkamp2022codegen,
  title={Codegen: An open large language model for code with multi-turn program synthesis},
  author={Nijkamp, Erik and Pang, Bo and Hayashi, Hiroaki and Tu, Lifu and Wang, Huan and Zhou, Yingbo and Savarese, Silvio and Xiong, Caiming},
  journal={arXiv preprint arXiv:2203.13474},
  year={2022}
}

@article{chen2021evaluating,
  title={Evaluating large language models trained on code},
  author={Mark Chen and Jerry Tworek and Heewoo Jun and Qiming Yuan and Henrique Ponde de Oliveira Pinto and Jared Kaplan and Harri Edwards and Yuri Burda and Nicholas Joseph and Greg Brockman and Alex Ray and Raul Puri and Gretchen Krueger and Michael Petrov and Heidy Khlaaf and Girish Sastry and Pamela Mishkin and Brooke Chan and Scott Gray and Nick Ryder and Mikhail Pavlov and Alethea Power and Lukasz Kaiser and Mohammad Bavarian and Clemens Winter and Philippe Tillet and Felipe Petroski Such and Dave Cummings and Matthias Plappert and Fotios Chantzis and Elizabeth Barnes and Ariel Herbert-Voss and William Hebgen Guss and Alex Nichol and Alex Paino and Nikolas Tezak and Jie Tang and Igor Babuschkin and Suchir Balaji and Shantanu Jain and William Saunders and Christopher Hesse and Andrew N. Carr and Jan Leike and Josh Achiam and Vedant Misra and Evan Morikawa and Alec Radford and Matthew Knight and Miles Brundage and Mira Murati and Katie Mayer and Peter Welinder and Bob McGrew and Dario Amodei and Sam McCandlish and Ilya Sutskever and Wojciech Zaremba},
  journal={arXiv preprint arXiv:2107.03374},
  year={2021}
}

@misc{wei2022emergentabilitieslargelanguage,
      title={Emergent Abilities of Large Language Models}, 
      author={Jason Wei and Yi Tay and Rishi Bommasani and Colin Raffel and Barret Zoph and Sebastian Borgeaud and Dani Yogatama and Maarten Bosma and Denny Zhou and Donald Metzler and Ed H. Chi and Tatsunori Hashimoto and Oriol Vinyals and Percy Liang and Jeff Dean and William Fedus},
      year={2022},
      eprint={2206.07682},
      archivePrefix={arXiv},
      primaryClass={cs.CL},
      url={https://arxiv.org/abs/2206.07682}, 
}

@misc{wei2023chainofthoughtpromptingelicitsreasoning,
      title={Chain-of-Thought Prompting Elicits Reasoning in Large Language Models}, 
      author={Jason Wei and Xuezhi Wang and Dale Schuurmans and Maarten Bosma and Brian Ichter and Fei Xia and Ed Chi and Quoc Le and Denny Zhou},
      year={2023},
      eprint={2201.11903},
      archivePrefix={arXiv},
      primaryClass={cs.CL},
      url={https://arxiv.org/abs/2201.11903}, 
}

@misc{pope2022efficientlyscalingtransformerinference,
      title={Efficiently Scaling Transformer Inference}, 
      author={Reiner Pope and Sholto Douglas and Aakanksha Chowdhery and Jacob Devlin and James Bradbury and Anselm Levskaya and Jonathan Heek and Kefan Xiao and Shivani Agrawal and Jeff Dean},
      year={2022},
      eprint={2211.05102},
      archivePrefix={arXiv},
      primaryClass={cs.LG},
      url={https://arxiv.org/abs/2211.05102}, 
}

@inproceedings{tpuv4,
author = {Jouppi, Norm and Kurian, George and Li, Sheng and Ma, Peter and Nagarajan, Rahul and Nai, Lifeng and Patil, Nishant and Subramanian, Suvinay and Swing, Andy and Towles, Brian and Young, Clifford and Zhou, Xiang and Zhou, Zongwei and Patterson, David A},
title = {TPU v4: An Optically Reconfigurable Supercomputer for Machine Learning with Hardware Support for Embeddings},
year = {2023},
isbn = {9798400700958},
publisher = {Association for Computing Machinery},
address = {New York, NY, USA},
url = {https://doi.org/10.1145/3579371.3589350},
doi = {10.1145/3579371.3589350},
booktitle = {Proceedings of the 50th Annual International Symposium on Computer Architecture},
articleno = {82},
numpages = {14},
location = {Orlando, FL, USA},
series = {ISCA '23}
}

@inproceedings{sun2024llumnix,
  title={Llumnix: Dynamic scheduling for large language model serving},
  author={Sun, Biao and Huang, Ziming and Zhao, Hanyu and Xiao, Wencong and Zhang, Xinyi and Li, Yong and Lin, Wei},
  booktitle={18th USENIX Symposium on Operating Systems Design and Implementation (OSDI 24)},
  pages={173--191},
  year={2024}
}

@inproceedings{gu2025pim,
  title={PIM is all you need: A CXL-enabled GPU-free system for large language model inference},
  author={Gu, Yufeng and Khadem, Alireza and Umesh, Sumanth and Liang, Ning and Servot, Xavier and Mutlu, Onur and Iyer, Ravi and Das, Reetuparna},
  booktitle={Proceedings of the 30th ACM International Conference on Architectural Support for Programming Languages and Operating Systems, Volume 2},
  pages={862--881},
  year={2025}
}

@misc{priority-queue,
    author = {Wikipedia},
    title = {Priority queue.},
    howpublished = {\url{https://en.wikipedia.org/wiki/Priority_queue}},
    year={2025}
}

@inproceedings{pan2025stratum,
  title={Stratum: System-Hardware Co-Design with Tiered Monolithic 3D-Stackable DRAM for Efficient MoE Serving},
  author={Yue Pan and
                  Zihan Xia and
                  Po{-}Kai Hsu and
                  Lanxiang Hu and
                  Hyungyo Kim and
                  Janak Sharda and
                  Minxuan Zhou and
                  Nam Sung Kim and
                  Shimeng Yu and
                  Tajana Rosing and
                  Mingu Kang},
  booktitle={Proceedings of the 58th IEEE/ACM International Symposium on Microarchitecture{\textregistered}},
  pages={1--17},
  year={2025}
}

@inproceedings{lee2021hardware,
  title={Hardware architecture and software stack for PIM based on commercial DRAM technology: Industrial product},
  author={Sukhan Lee and
                  Shinhaeng Kang and
                  Jaehoon Lee and
                  Hyeonsu Kim and
                  Eojin Lee and
                  Seungwoo Seo and
                  Hosang Yoon and
                  Seungwon Lee and
                  Kyounghwan Lim and
                  Hyunsung Shin and
                  Jinhyun Kim and
                  Seongil O and
                  Anand Iyer and
                  David Wang and
                  Kyomin Sohn and
                  Nam Sung Kim},
  booktitle={2021 ACM/IEEE 48th Annual International Symposium on Computer Architecture (ISCA)},
  pages={43--56},
  year={2021},
  organization={IEEE}
}

@inproceedings{he2020newton,
  title={Newton: A DRAM-maker’s accelerator-in-memory (AiM) architecture for machine learning},
  author={He, Mingxuan and Song, Choungki and Kim, Ilkon and Jeong, Chunseok and Kim, Seho and Park, Il and Thottethodi, Mithuna and Vijaykumar, TN},
  booktitle={2020 53rd Annual IEEE/ACM International Symposium on Microarchitecture (MICRO)},
  pages={372--385},
  year={2020},
  organization={IEEE}
}

@article{kim2024breakthrough,
  title={The breakthrough memory solutions for improved performance on llm inference},
  author={Byeongho Kim and
                  Sanghoon Cha and
                  Sangsoo Park and
                  Jieun Lee and
                  Sukhan Lee and
                  Shinhaeng Kang and
                  Jinin So and
                  Kyungsoo Kim and
                  Jin Jung and
                  Jong{-}Geon Lee and
                  Sunjung Lee and
                  Yoonah Paik and
                  Hyeonsu Kim and
                  Jin{-}Seong Kim and
                  Won{-}Jo Lee and
                  Yuhwan Ro and
                  Yeongon Cho and
                  Jin Hyun Kim and
                  Joon{-}Ho Song and
                  Jaehoon Yu and
                  Seungwon Lee and
                  Jeonghyeon Cho and
                  Kyomin Sohn},
  journal={IEEE Micro},
  volume={44},
  number={3},
  pages={40--48},
  year={2024},
  publisher={IEEE}
}

@inproceedings{lee20221ynm,
  title={A 1ynm 1.25 V 8Gb, 16Gb/s/pin GDDR6-based accelerator-in-memory supporting 1TFLOPS MAC operation and various activation functions for deep-learning applications},
  author={Seong Ju Lee and
                  Kyu{-}Young Kim and
                  Sanghoon Oh and
                  Joonhong Park and
                  Gimoon Hong and
                  Dong Yoon Ka and
                  Kyu{-}Dong Hwang and
                  Jeongje Park and
                  Kyeong Pil Kang and
                  Jungyeon Kim and
                  Junyeol Jeon and
                  Nahsung Kim and
                  Yongkee Kwon and
                  Kornijcuk Vladimir and
                  Woojae Shin and
                  Jongsoon Won and
                  Minkyu Lee and
                  Hyunha Joo and
                  Haerang Choi and
                  Jaewook Lee and
                  Donguc Ko and
                  Younggun Jun and
                  Keewon Cho and
                  Ilwoong Kim and
                  Choungki Song and
                  Chunseok Jeong and
                  Dae{-}Han Kwon and
                  Jieun Jang and
                  Il Park and
                  Junhyun Chun and
                  Joohwan Cho},
  booktitle={2022 IEEE International Solid-State Circuits Conference (ISSCC)},
  volume={65},
  pages={1--3},
  year={2022},
  organization={IEEE}
}

@inproceedings{oprins20203d,
  title={3D wafer-to-wafer bonding thermal resistance comparison: Hybrid Cu/dielectric bonding versus dielectric via-last bonding},
  author={Oprins, Herman and Cherman, Vladimir and Webers, Tomas and Kim, Soon-Wook and de Vos, Joeri and Van der Plas, Geert and Beyne, Eric},
  booktitle={2020 19th IEEE Intersociety Conference on Thermal and Thermomechanical Phenomena in Electronic Systems (ITherm)},
  pages={219--228},
  year={2020},
  organization={IEEE}
}

@article{ren2020thermal,
  title={Thermal TSV optimization and hierarchical floorplanning for 3-D integrated circuits},
  author={Ren, Zongqing and Alqahtani, Ayed and Bagherzadeh, Nader and Lee, Jaeho},
  journal={IEEE Transactions on Components, Packaging and Manufacturing Technology},
  volume={10},
  number={4},
  pages={599--610},
  year={2020},
  publisher={IEEE}
}

@inproceedings{han20222,
  title={From 2.5 D to 3D chiplet systems: Investigation of thermal implications with HotSpot 7.0},
  author={Han, Jun-Han and Guo, Xinfei and Skadron, Kevin and Stan, Mircea R},
  booktitle={2022 21st IEEE Intersociety Conference on Thermal and Thermomechanical Phenomena in Electronic Systems (iTherm)},
  pages={1--6},
  year={2022},
  organization={IEEE}
}

@inproceedings{yue20253d,
  title={3D-PATH: A Hierarchy LUT Processing-in-memory Accelerator with Thermal-aware Hybrid Bonding Integration},
  author={Yue, Zhiheng and Wang, Yang and Li, Chao and Wei, Shaojun and Hu, Yang and Yin, Shouyi},
  booktitle={Proceedings of the 58th IEEE/ACM International Symposium on Microarchitecture},
  pages={78--93},
  year={2025}
}

@misc{waferllm0,
    author = {He, Congjie and Huang, Yeqi and Mu, Pei and Miao, Ziming and Xue, Jilong and Ma, Lingxiao and Yang, Fan and Mai, Luo},
    title = {WaferLLM's Decoding Implementation.},
    howpublished = {\url{https://github.com/MeshInfra/WaferLLM/blob/7122d0fac527a85b440804f930e0c43b5df88102/Decode/WSE-3/}},
    year={2025}
}

@misc{waferllm1,
    author = {He, Congjie and Huang, Yeqi and Mu, Pei and Miao, Ziming and Xue, Jilong and Ma, Lingxiao and Yang, Fan and Mai, Luo},
    title = {K Cache Shape Definition in WaferLLM's Decoding Implementation.},
    howpublished = {\url{https://github.com/MeshInfra/WaferLLM/blob/7122d0fac527a85b440804f930e0c43b5df88102/Decode/WSE-3/src/decode.csl#L239}},
    year={2025}
}

@misc{waferllm2,
    author = {He, Congjie and Huang, Yeqi and Mu, Pei and Miao, Ziming and Xue, Jilong and Ma, Lingxiao and Yang, Fan and Mai, Luo},
    title = {V Cache Shape Definition in WaferLLM's Decoding Implementation.},
    howpublished = {\url{https://github.com/MeshInfra/WaferLLM/blob/7122d0fac527a85b440804f930e0c43b5df88102/Decode/WSE-3/src/decode.csl#L243}},
    year={2025}
}

@misc{waferllm3,
    author = {He, Congjie and Huang, Yeqi and Mu, Pei and Miao, Ziming and Xue, Jilong and Ma, Lingxiao and Yang, Fan and Mai, Luo},
    title = {The use of K Cache in WaferLLM's Decoding Implementation.},
    howpublished = {\url{https://github.com/MeshInfra/WaferLLM/blob/7122d0fac527a85b440804f930e0c43b5df88102/Decode/WSE-3/src/decode.csl#L556}},
    year={2025}
}

@misc{waferllm4,
    author = {He, Congjie and Huang, Yeqi and Mu, Pei and Miao, Ziming and Xue, Jilong and Ma, Lingxiao and Yang, Fan and Mai, Luo},
    title = {The use of V Cache in WaferLLM's Decoding Implementation.},
    howpublished = {\url{https://github.com/MeshInfra/WaferLLM/blob/7122d0fac527a85b440804f930e0c43b5df88102/Decode/WSE-3/src/decode.csl#L617}},
    year={2025}
}

@misc{waferllm5,
    author = {He, Congjie and Huang, Yeqi and Mu, Pei and Miao, Ziming and Xue, Jilong and Ma, Lingxiao and Yang, Fan and Mai, Luo},
    title = {The Definition of Fixed Context Length before WaferLLM's Decoding Execution.},
    howpublished = {\url{https://github.com/MeshInfra/WaferLLM/blob/7122d0fac527a85b440804f930e0c43b5df88102/Decode/WSE-3/run_wse3.sh#L48}},
    year={2025}
}

@misc{waferllm6,
    author = {He, Congjie and Huang, Yeqi and Mu, Pei and Miao, Ziming and Xue, Jilong and Ma, Lingxiao and Yang, Fan and Mai, Luo},
    title = {Setting Fixed Context Length when calling WaferLLM's Decoding Execution.},
    howpublished = {\url{https://github.com/MeshInfra/WaferLLM/blob/7122d0fac527a85b440804f930e0c43b5df88102/Decode/WSE-3/run_wse3.sh#L71}},
    year={2025}
}

@misc{waferllm7,
    author = {He, Congjie and Huang, Yeqi and Mu, Pei and Miao, Ziming and Xue, Jilong and Ma, Lingxiao and Yang, Fan and Mai, Luo},
    title = {WaferLLM's Decoding Entry and Operator Arrangement.},
    howpublished = {\url{https://github.com/MeshInfra/WaferLLM/blob/7122d0fac527a85b440804f930e0c43b5df88102/Decode/WSE-3/src/decode.csl#L765}},
    year={2025}
}

@misc{waferllm8,
    author = {He, Congjie and Huang, Yeqi and Mu, Pei and Miao, Ziming and Xue, Jilong and Ma, Lingxiao and Yang, Fan and Mai, Luo},
    title = {WaferLLM's Shift Operator Implementation.},
    howpublished = {\url{https://github.com/MeshInfra/WaferLLM/tree/7122d0fac527a85b440804f930e0c43b5df88102/Shift}},
    year={2025}
}

@misc{waferllm9,
    author = {He, Congjie and Huang, Yeqi and Mu, Pei and Miao, Ziming and Xue, Jilong and Ma, Lingxiao and Yang, Fan and Mai, Luo},
    title = {Explicit Softmax between two Attention GEMMs in WaferLLM's Decoding Implementation.},
    howpublished = {\url{https://github.com/MeshInfra/WaferLLM/blob/7122d0fac527a85b440804f930e0c43b5df88102/Decode/WSE-3/src/decode.csl#L783}},
    year={2025}
}

\end{document}